\documentclass{ws-ijmpe}
\usepackage[super,compress]{cite}
\usepackage[breaklinks]{hyperref}
\hypersetup{colorlinks,urlcolor=black,citecolor=red,linkcolor=green,filecolor=black}
\usepackage{breakurl}

\begin{document}

\markboth{Xiang-Quan Deng and Shan-Gui Zhou}{Study of U isotopes with MDC-RMF model}

\catchline{}{}{}{}{}

\title{Ground state and fission properties of even-$A$ uranium isotopes from
       multidimensionally-constrained relativistic mean field model}

\author{Xiang-Quan Deng$^{1,2}$ and Shan-Gui Zhou$^{1,2,3,4,5,*}$}

\address{%
$^1$CAS Key Laboratory of Theoretical Physics, Institute of Theoretical Physics,
    Chinese Academy of Sciences, Beijing 100190, China \\
$^2$School of Physical Sciences, University of Chinese Academy of Sciences,
    Beijing 100049, China \\
$^3$School of Nuclear Science and Technology, University of Chinese Academy of Sciences,
    Beijing 100049, China \\
$^4$Synergetic Innovation Center for Quantum Effects and Application,
    Hunan Normal University, Changsha, 410081, China \\
$^5$Peng Huanwu Collaborative Center for Research and Education,
    Beihang University, Beijing 100191, China \\
$^*$sgzhou@itp.ac.cn}

\maketitle

\begin{history}
\received{Day Month Year}
\revised{Day Month Year}
\end{history}

\begin{abstract}
The multidimensionally-constrained covariant density functional theories
(MDC-CDFTs) have been developed to study the influence of
octupole and triaxial deformations on the ground state and fission properties.
In this paper, we present a brief review of the applications of MDC-CDFTs and
discuss the results of a systematical study of even-$A$ uranium isotopes
with the MDC-RMF model which is one of MDC-CDFTs
with pairing correlations treated by using the BCS approach.
We examine in detail the two-dimensional potential energy surfaces
$E(\beta_{20},\beta_{30})$ of these U isotopes and discuss the ground state and
fission properties as well as third and fourth minima on the potential
energy surfaces.
The emphasis is put on the effects of octupole and triaxial deformations.
\end{abstract}




\section{Introduction}

Atomic nuclei are quantum many-body systems consisting of nucleons,
i.e., protons and neutrons.
The ground states of nuclei are characterized by structure properties
such as the spin and parity $I^\pi$, mass excess or binding energy, size and shape,
which are governed
by both the nucleon-nucleon interaction and many-body features
from which various nuclear phenomena emerge
\cite{Bohr1998_Nucl_Structure_1,Bohr1998_Nucl_Structure_2,Ring1980}.

In the intrinsic frame, many nuclear shapes may appear and manifest themselves
in the low-lying collective spectra:
The rotational spectrum $E(I) \sim I(I+1)$ corresponds to an axial quadrupole deformation
\cite{Bohr1953_PR089-316,Bohr1953_PR090-717};
a static octupole deformation results in the parity doublet bands
\cite{Alder1956_RMP28-432,Alder1958_RMP30-353,Bohr1957_NP4-529,Bohr1958_NP9-687,Bohr1998_Nucl_Structure_2};
the triaxial quadrupole deformation is characterized by the chiral doublet bands
\cite{Frauendorf1997_NPA617-131,Meng2006_PRC73-037303}
or the wobbling motion \cite{Bohr1998_Nucl_Structure_2} in certain nuclei.
More exotic intrinsic nuclear shapes have been explored extensively, e.g.,
the hyperdeformed shapes, see Ref.~\citenum{Zhao2015_PRC91-014321} and references therein,
the rod shapes \cite{Zhao2015_PRL115-022501},
the tetrahedral or octahedral shapes \cite{Dudek2006_PRL97-072501,Chen2010_NPA834-378c,Zhao2017_PRC95-014320},
the triangle shape \cite{Wang2022_CTP74-015303}
and
the toroidal or ring shapes \cite{Wong1973_APNY77-279,Cao2019_PRC99-014606}.



Nuclear fission---the large amplitude collective motion---can be described as
the evolution of nuclear shape in a multidimensional deformation space
\cite{Bohr1939_PR056-426,Strutinsky1967_NPA95-420,Strutinsky1968_NPA122-1,
Moeller1973_IAEA-SM-174-202,Lu2012_PRC85-011301R}.
The potential energy surface (PES) in such a deformation space is crucial for
the study of nuclear fission
\cite{Brack1972_RMP44-320,Abusara2010_PRC82-044303,Abusara2012_PRC85-024314,
Lu2012_PRC85-011301R,
Lu2014_PRC89-014323,Zhao2015_PRC91-014321,Schunck2016_RPP79-116301,
Schmidt2018_RPP81-106301,Schunck2022_PPNP125-103963}.
In particular, the characteristics of fission barrier, e.g., the height
and width, are important inputs for theoretical models of
nuclear fission \cite{Sheikh2009_PRC80-011302R,Kowal2010_PRC82-014303,
Eichler2015_ApJ808-30,
Sonzogni2016_PRL116-132502,Agbemava2017_PRC95-054324,
Brown2018_NDS148-1,Ge2020_EPJWoC239-09001,Plompen2020_EPJA56-181}.
For instance, in the statistical models for calculating the survival probability
of hot compound superheavy nuclei (SHN),
the competition between the neutron evaporation and fission is mainly
determined by the neutron separation energy and fission barrier height
\cite{Zubov2002_PRC65-024308,Xia2011_SciChinaPMA54S1-109,Wang2012_PRC85-041601R,
Adamian2021_EPJA57-89,Deng2023_PRC107-014616}.
It has been revealed that,
besides the axial quadrupole deformation
which is the most important shape degree of freedom in describing nuclear fission,
the nonaxial and reflection asymmetric deformations are crucial as well
\cite{Moeller1973_IAEA-SM-174-202,Lu2012_PRC85-011301R}.

Many theoretical models have been developed and used to study the PESs and barrier heights,
including the macroscopic-microscopic (MM) models
\cite{Howard1980_ADNDT25-219,Myers1996_NPA601-141,Jachimowicz2012_PRC85-034305,
Staszczak2013_PRC87-024320,
Pomorski2013_PST154-014023,Zhong2014_CTP62-405,
Moeller2015_PRC91-024310,Pomorski2017_PS92-064006,Chai2018_ChinPhysC42-054101,
Jachimowicz2020_PRC101-014311,Zhu2020_CTP72-105301},
the extended Thomas-Fermi plus Strutinsky intergral (ETFSI) method
\cite{Mamdouh2001_NPA679-337}
and the density functional theories (DFTs)
\cite{Delaroche2006_NPA771-103,Goriely2009_PRC79-024612,
Giuliani2013_PRC88-054325,
Rodriguez-Guzman2014_PRC89-054310,
Zhou2016_PS91-063008,
Giuliani2018_PRC97-034323,Ling2020_EPJA56-180,Taninah2020_PRC102-054330}.
The multidimensionally-constrained (MDC) covariant density functional theories (CDFTs)
have been developed for the study of ground state and fission properties as well as the PESs
\cite{Lu2014_PRC89-014323,Zhou2016_PS91-063008,Zhao2017_PRC95-014320}.
In MDC-CDFTs, both nonaxial and reflection asymmetric deformations are
considered self-consistently.
The MDC-CDFTs include the MDC relativistic mean field (MDC-RMF) model
in which the pairing correlations are treated by using the BCS approach
and the MDC relativistic Hartree-Bogoliubov (MDC-RHB) model.

In the present work, by using the MDC-RMF model,
we study systematically the PESs of even-$A$ U isotopes from the two-proton drip line
up to the two-neutron drip line.
We study the ground state properties and the primary barrier heights
and examine the effects of octupole and triaxial deformations.
The third and fourth minima appearing on the PESs of some U isotopes are also discussed.

The paper is organized as follows.
The applications of the MDC-CDFTs is briefly reviewed in Sec.~\ref{sec2}.
In Sec.~\ref{sec3}, we present numerical results of the PESs and
discuss in detail the ground state and fission properties of even-$A$ U isotopes.
A summary of this work is given in Sec.~\ref{sec4}.


\section{Brief Review of MDC-CDFTs}\label{sec2}

The CDFT has been very successful in self-consistent descriptions
of atomic nuclei throughout almost the whole chart of nuclides
\cite{Serot1986_ANP16-1,Reinhard1989_RPP52-439,Ring1996_PPNP37-193,Vretenar2005_PR409-101,
Meng2006_PPNP57-470,Niksic2011_PPNP66-519,Meng2013_FrontPhys8-55,Liang2015_PR570-1,
Meng2015_JPG42-093101,Zhou2016_PS91-063008,
Meng2016_RDFNS,Shen2019_PPNP109-103713}.
In the CDFT, one usually derives the equations of motion of nucleons
from the Lagrangian incorporating the nucleon fields and
the interaction between nucleons which is realized
either via meson exchanges (ME) or through point couplings (PC).

In the MDC-CDFTs, the equations of motion for nucleons are solved
with the basis expansion method.
For axially deformed nuclei with the reflection asymmetry,
a two-center basis would be more appropriate;
e.g.,
the reflection asymmetric RMF (RAS-RMF) model \cite{Geng2007_CPL24-1865} has been developed
in a two-center harmonic-oscillator basis
and used to study octupole deformations in
$^{226}$Ra \cite{Geng2007_CPL24-1865} and
Sm \cite{Zhang2010_PRC81-034302},
Ba \cite{Wei2010_ChinPhysC34-1094},
Th \cite{Guo2010_PRC82-047301} and
Dy \cite{Qiu2022_PRC106-034301} isotopes.
However, in the MDC-CDFTs, the reflection symmetric, axially deformed
harmonic oscillator (ADHO) basis
\cite{Gambhir1990_APNY198-132} was adopted for convenience.
In Ref.~\citenum{Gambhir1990_APNY198-132}, the nuclei in question were assumed
to be spherical or axially deformed with the reflection symmetry, i.e.,
in the following multipole expansion of the nuclear surface,
only even $\lambda$ is considered and $\mu = 0$,
\begin{equation}
 R ( \theta, \varphi ) =
 R_0 \left[  1 +
           \sum_{\lambda=2}^{\infty} \sum_{\mu=-\lambda}^\lambda
            \beta_{\lambda \mu}^* Y_{\lambda \mu} ( \theta, \varphi )
     \right] .
 \label{Eq:SurfaceDeformation}
\end{equation}
In the MDC-CDFTs, the $V_4$ symmetry was imposed
when solving the equations of motion for nucleons.
Thus all deformations $\beta_{\lambda\mu}$ with even $\mu$,
including nonaxial and reflection asymmetric deformations,
are included self-consistently.
The pairing correlations are treated by using the BCS method in the MDC-RMF model
\cite{Lu2014_PRC89-014323}
and by implementing the Bogoliubov transformation in the MDC-RHB model
\cite{Zhao2017_PRC95-014320}.
For the details of the MDC-CDFTs, the readers are referred to
Refs.~\citenum{Lu2014_PRC89-014323,Zhao2017_PRC95-014320,Zhou2016_PS91-063008}.

The MDC-CDFTs have been extensively used to study normal and hypernuclei,
with emphasis on the properties concerning various deformation effects
on nuclear properties including the ground state properties, PESs,
superdeformed and hyperdeformed states and fission properties.
Next we present a brief review of the applications and further developments of the MDC-CDFTs.
\begin{description}

\item[Fission barriers of actinide nuclei]

The MDC-CDFTs have been applied to the study of fission barriers of actinide nuclei
\cite{Lu2012_PRC85-011301R,Lu2014_PRC89-014323}.
The typical double-humped fission barriers in actinides were well reproduced.
Besides the lowering effects of the triaxial distortion on the first barrier and
the large influence of the octupole deformation on the second barrier,
it was found that the triaxial deformation also lowers the second barrier considerably.
When both triaxial and octupole deformations are included,
the calculated fission barriers heights agree satisfactorily
with the available empirical values.

\item[Third minima in PESs of light actinides]

The PESs of light actinides were carefully examined and the third minima
were investigated \cite{Zhao2015_PRC91-014321}.
The origin of these minima, corresponding to hyperdeformed states,
has been attributed to
the $Z=90$ proton shell gap at very large deformations.

\item[Shapes and PESs of superheavy nuclei]

One- and two-dimensional PESs have been obtained for $^{270}$Hs---a deformed doubly magic SHN
\cite{Meng2020_SciChinaPMA63-212011}.
The influences of the nonaxial and reflection asymmetric distortions on
the fission barrier and fission pathway have been investigated:
When the axial symmetry is imposed,
the reflection symmetric and asymmetric fission paths both show a double-humped structure
and the latter is energetically favored;
when nonaxial shapes are allowed,
the reflection symmetric fission pathway becomes favorable.
The higher-order deformation effects
on the ground state properties of SHN of and near $^{270}$Hs have been studied
\cite{Wang2022_ChinPhysC46-024107}.
It was found that among higher-order deformations $\beta_\lambda$ ($\lambda=4,6,8 $ and $10$),
$\beta_6$ has the greatest impact.

\item[Non-axial octupole $Y_{32}$ correlations]

The nonaxial reflection-asymmetric $\beta_{32}$ shape in transfermium nuclei
with $N=150$ have been studied \cite{Zhao2012_PRC86-057304}.
It was found that in these nuclei, the origin of the $Y_{32}$ correlations
is mainly from a pair of neutron orbitals,
$[734]9/2 (\nu j_{15/2})$ and $[622]5/2 (\nu g_{9/2})$,
and a pair of proton orbitals, $[521]3/2 (\pi f_{7/2})$ and $[633]7/2 (\pi i_{13/2})$.
The tetrahedral shapes in neutron-rich even-even Zr isotopes have been investigated
\cite{Zhao2017_PRC95-014320}.
The tetrahedral ground states are mainly caused
by large shell gaps around $Z=40$ and $N=70$.

\item[Axial octupole $Y_{30}$ correlations]

The MDC-CDFTs have been used to investigate the coexistence of chirality and
octupole correlations in
$^{76}$Br \cite{Xu2022_PLB833-137287} and
$^{78}$Br \cite{Liu2016_PRL116-112501,Wang2019_PLB792-454};
the latter is the first example of chiral geometry in octupole soft nuclei.
The octupole correlations in
$^{123,125}$Ba \cite{Chen2016_PRC94-021301R},
$^{73}$Br \cite{Wang2022_PRC105-044316}
and $^{71}$Ge \cite{Wang2022_PRC106-L011303}
have been revealed with the MDC-CDFTs.

\item[Structure of hypernuclei]

The structure of hypernuclei has been studied by
using the MDC-CDFTs
\cite{Lu2011_PRC84-014328,Lu2014_PRC89-044307,Rong2020_PLB807-135533,Rong2021_PRC104-054321,
Chen2021_SciChinaPMA64-282011,Sun2022_ChinPhysC46-074106}.
For brevity, we only briefly mention two works concerning the deformation effects.
In the study of the shapes of light normal and $\Lambda$ hypernuclei,
it was found that
the shape polarization effect of $\Lambda$ is so strong that
the shapes of some $\Lambda$ hypernuclei,
e.g., $^{13}_\Lambda$C, $^{23}_\Lambda$C, and $^{31}_\Lambda$Si,
are drastically different from the corresponding cores
\cite{Lu2011_PRC84-014328}.
The superdeformed (SD) states and corresponding SD $\Lambda$ hypernuclei of Ar isotopes
were examined.
A strong localization effect with a ring structure was predicted
in density distributions of the SD states,
resulting in a larger $\Lambda$ separation energy in SD states than in the ground states
\cite{Lu2014_PRC89-044307}.

\item[Fission dynamics and $\alpha$ decay]

Based on the PESs from the MDC-CDFTs, the dynamics of spontaneous and
induced fissions in actinide nuclei has been studied extensively
\cite{Zhao2015_PRC92-064315,Zhao2016_PRC93-044315,Zhao2019_PRC99-014618,Zhao2019_PRC99-054613,
Zhao2020_PRC102-054606,Zhao2020_PRC101-064605,Zhao2021_PRC104-044612,Ren2022_PRC105-044313,
Zhao2022_PRC105-054604,Zhao2022_PRC106-054609}.
For example, in Ref.~\citenum{Zhao2016_PRC93-044315},
the spontaneous fission dynamics of $^{264}$Fm and $^{250}$Fm was explored
and it was concluded that the inclusion of pairing correlations in the space of
collective coordinates favors axially symmetric shapes
along the dynamic pathway of the fissioning system.
Such dynamic studies tell us more about the role played by the various deformations
in fission than static studies focusing only on PESs.
Recently, microscopic investigations of half-lives for $\alpha$ decays in
$^{108}\mathrm{Xe}$ and $^{104}\mathrm{Te}$
\cite{Mercier2020_PRC102-011301} and $\ensuremath{\alpha}$
and $2\ensuremath{\alpha}$ decays in $^{212}\mathrm{Po}$ and $^{224}\mathrm{Ra}$
\cite{Mercier2021_PRL127-012501}
have also been performed based on the PESs calculated from the MDC-CDFTs.

\item[Angular momentum and parity projections]

The angular momentum projection (AMP) and parity projection (PP) have been
implemented in the MDC-RHB model to restore the rotational and parity symmetries which
are both broken in the mean-field level.
Such a projected-MDCRHB (p-MDCRHB) model was presented in Ref.~\citenum{Wang2022_CTP74-015303}.
With the p-MDCRHB model, one may study nuclear spectra corresponding to
exotic intrinsic shapes, e.g., triangle or tetrahedron.
In Ref.~\citenum{Rong2023_PLB-submitted}, an anatomy of octupole correlations
in $^{96}$Zr has been performed with the p-MDCRHB model.
It was found that the PESs of this nucleus are strongly dependent on
the angular momentum and parity and both triaxial and
octupole deformations should be included in order to give
a decent description of the structure of $^{96}$Zr.

\end{description}

\section{Results and Discussions}\label{sec3}

In this section, we systematically study the properties of
even-$A$ U isotopes with the MDC-RMF model.
Firstly, we present and discuss two-dimensional (2D) PESs of several selected U isotopes.
Secondly, we focus on the ground state properties of the isotopes and
show the density distribution profiles of several typical nuclei
with deformed ground state shapes.
Thirdly, for each isotope, we perform calculations in the vicinity of each saddle point
with the reflection symmetry and the axial symmetry broken.
The primary barrier heights with octupole and triaxial deformations are compared with
the available empirical values and/or calculation results of other models.
Finally, we discuss hyperdeformed third and even fourth minima on the PESs of
several U isotopes.

In our calculations, the effective interaction PC-PK1
\cite{Zhao2010_PRC82-054319} is used.
The truncation in the expansion of the large component of
the Dirac spinor $N_\mathrm{f}=20$
and that for the small component $N_\mathrm{g}=21$.
The quadrupole deformation parameter of the ADHO basis is
chosen to be half of the desired $\beta_{20}$ value.
The pairing correlations are treated with the BCS approach and
the separable pairing force of finite-range
\cite{Tian2006_CPL23-3226,Tian2009_PLB676-44,Tian2009_PRC79-064301} is used.
The pairing strength and effective range are taken as $G=1.1G_0$ with
$G_0=728$ MeV $\mathrm{fm^3}$ and $a=0.644$ fm.

\subsection{Two-dimensional PESs}

We calculate 2D PESs $E(\beta_{20},\beta_{30})$ of the U isotopes,
from the two-proton drip line $^{214}$U,
the lightest known uranium isotope \cite{Zhang2021_PRL126-152502},
to the two-neutron drip line $^{350}$U.
In these calculations the nuclei are assumed to be axially symmetric.
For each isotope, $\beta_{20}$ runs from $-0.20$ to 2.00 and
$\beta_{30}$ from 0.00 to 1.00, both with a step size of 0.05.
The PESs with negative $\beta_{30}$ value are obtained through
the relation $E(\beta_{20},\beta_{30})=E(\beta_{20},\lvert\beta_{30}\rvert )$.

\begin{figure*}[htbp]
\centering
\begin{tabular}{ccc}
\includegraphics[width=0.46\textwidth]{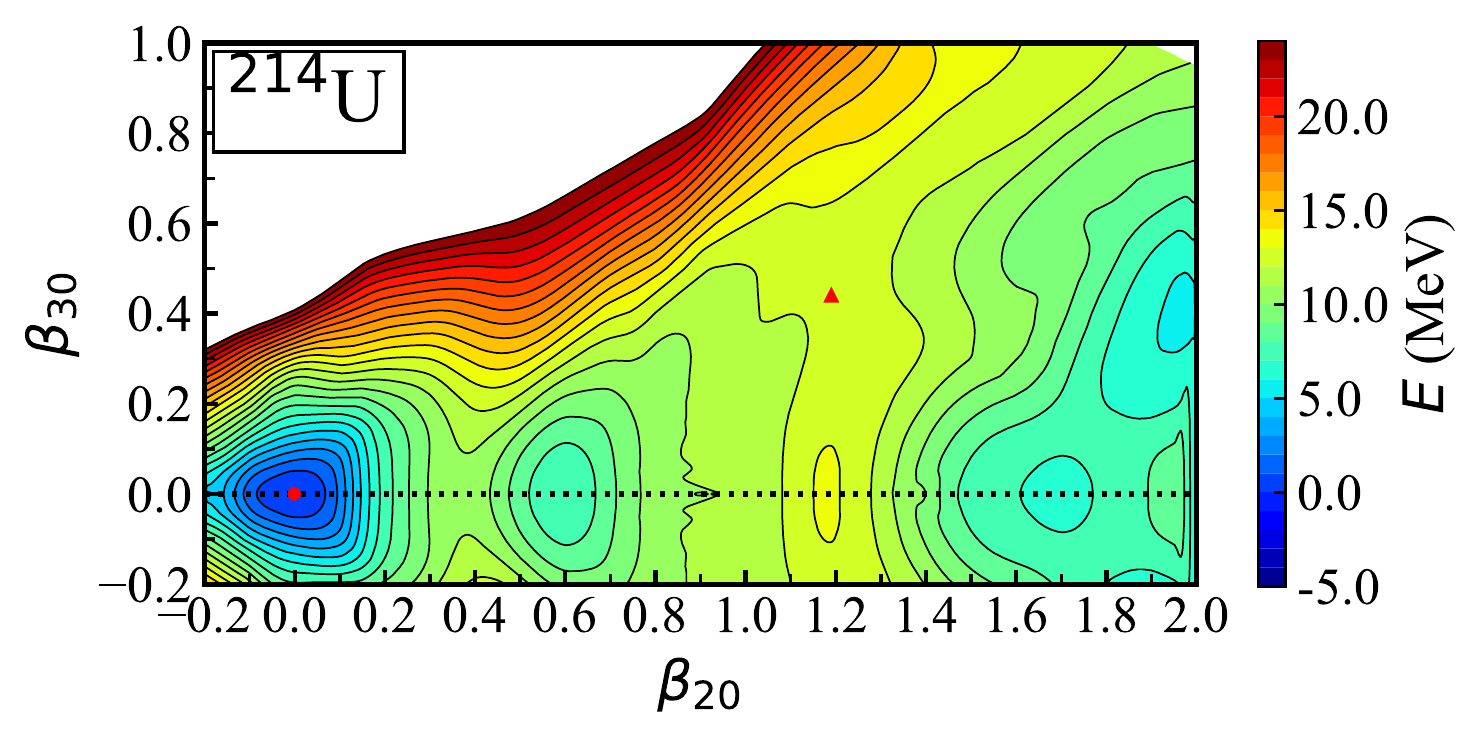}
\includegraphics[width=0.46\textwidth]{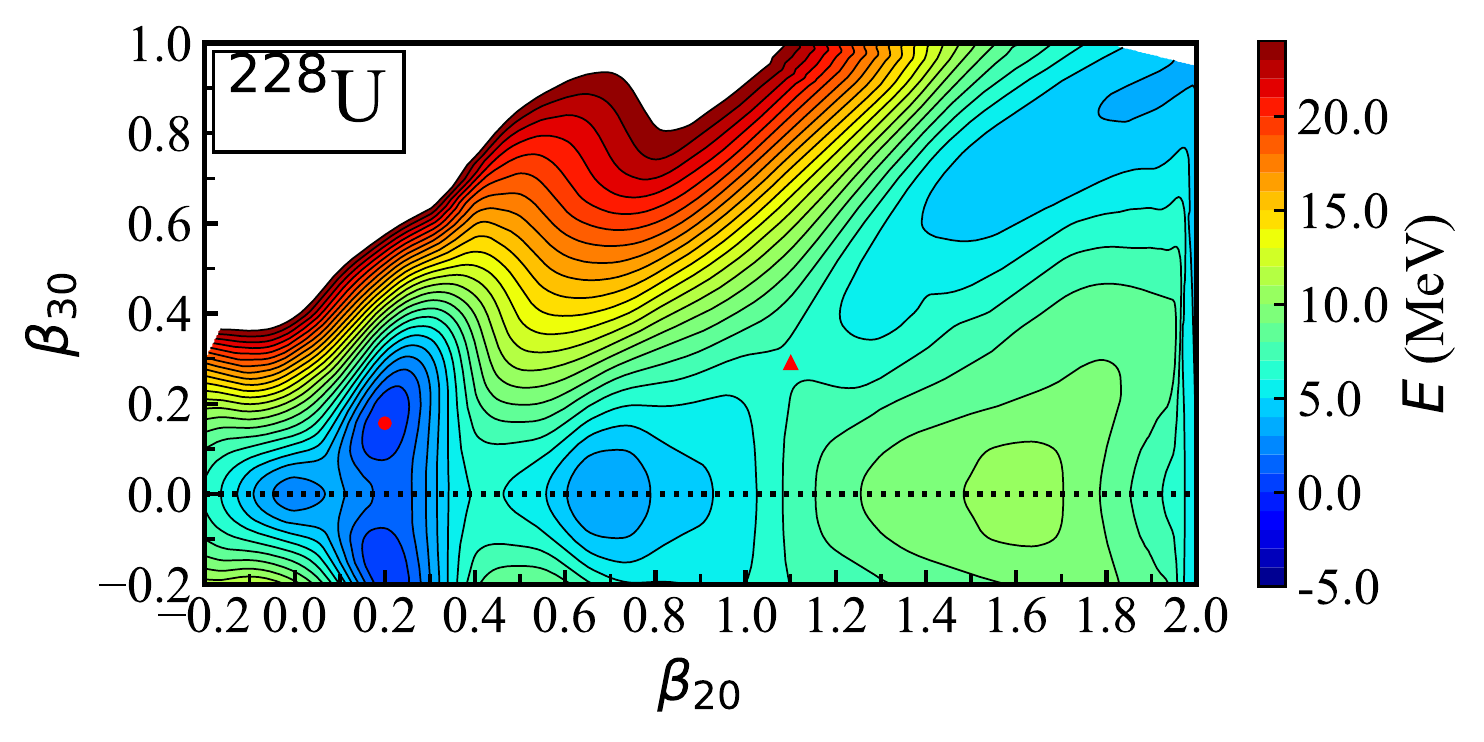}\\
\includegraphics[width=0.46\textwidth]{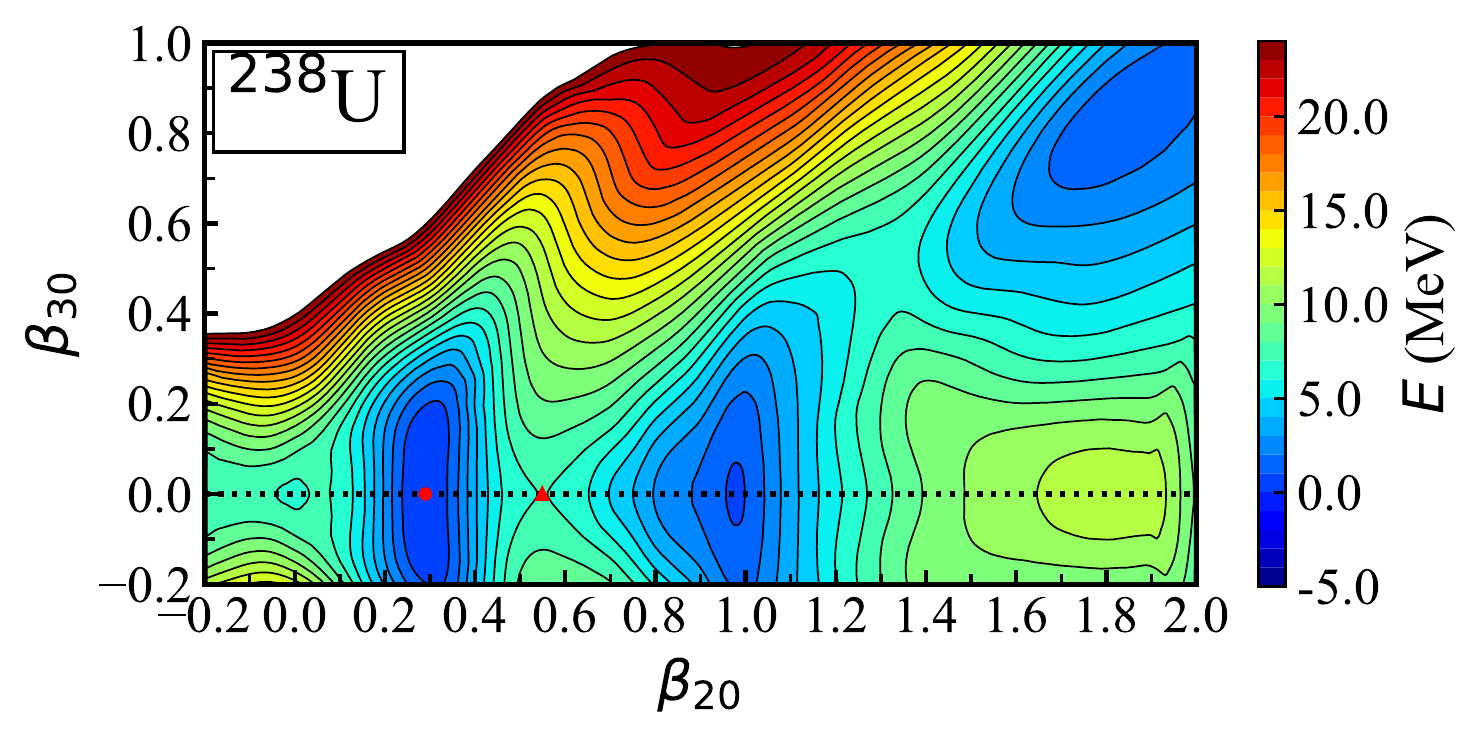}
\includegraphics[width=0.46\textwidth]{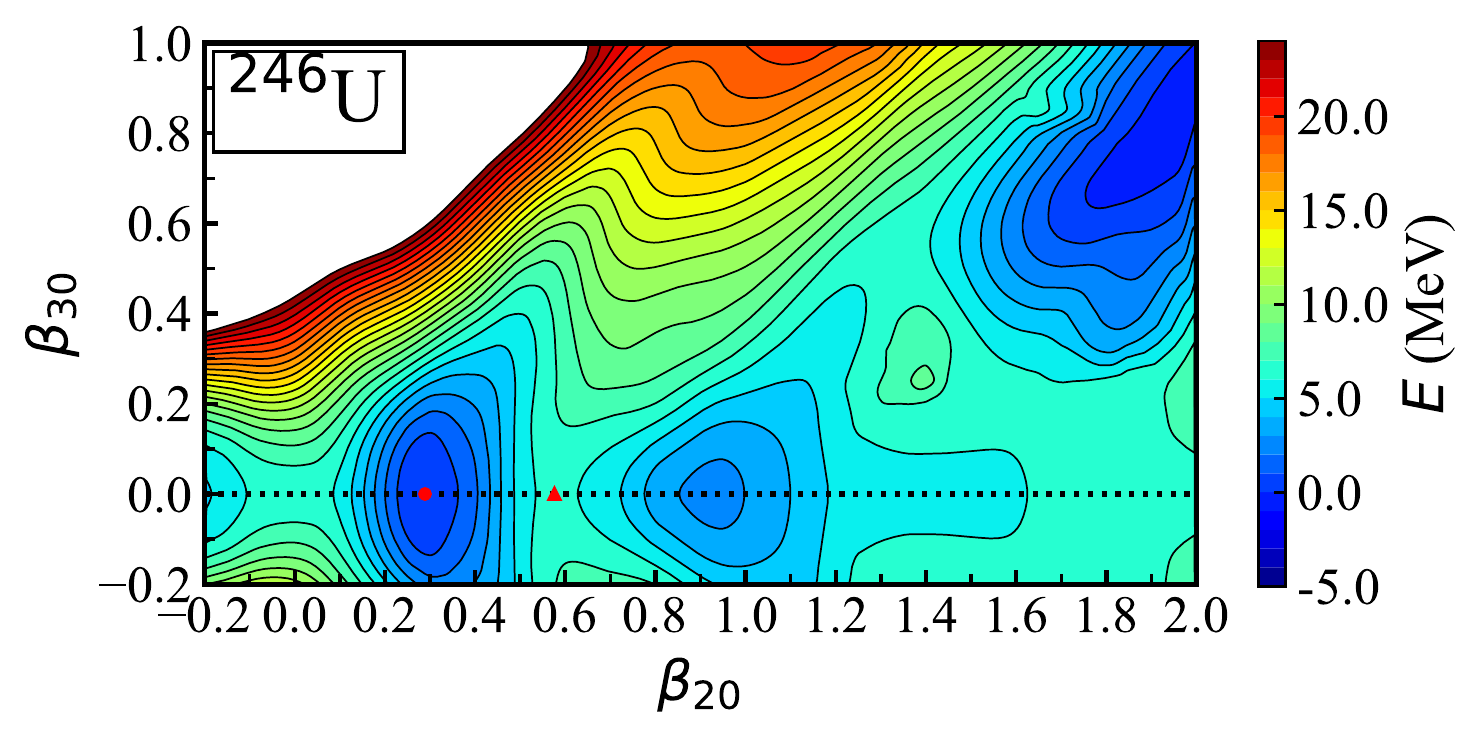}\\
\includegraphics[width=0.46\textwidth]{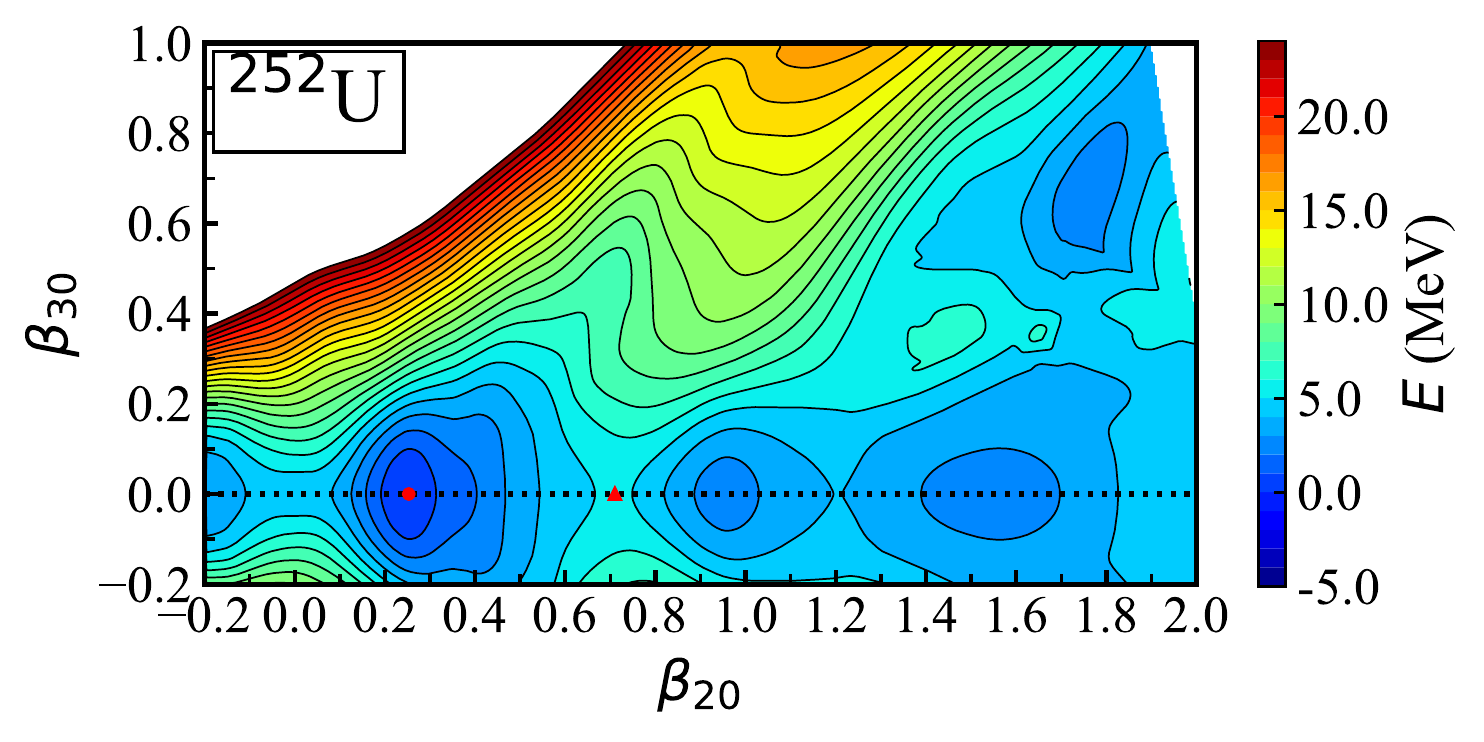}
\includegraphics[width=0.46\textwidth]{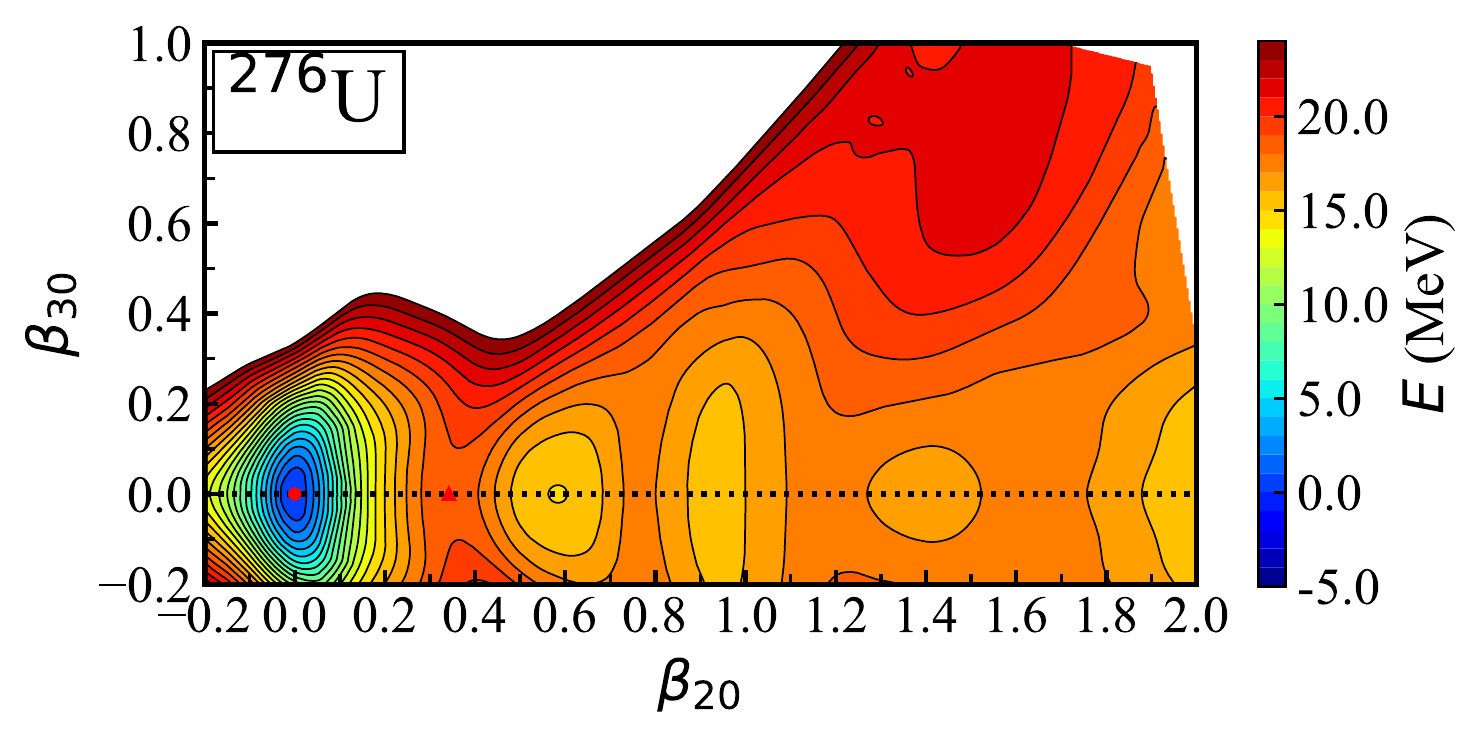}\\
\includegraphics[width=0.46\textwidth]{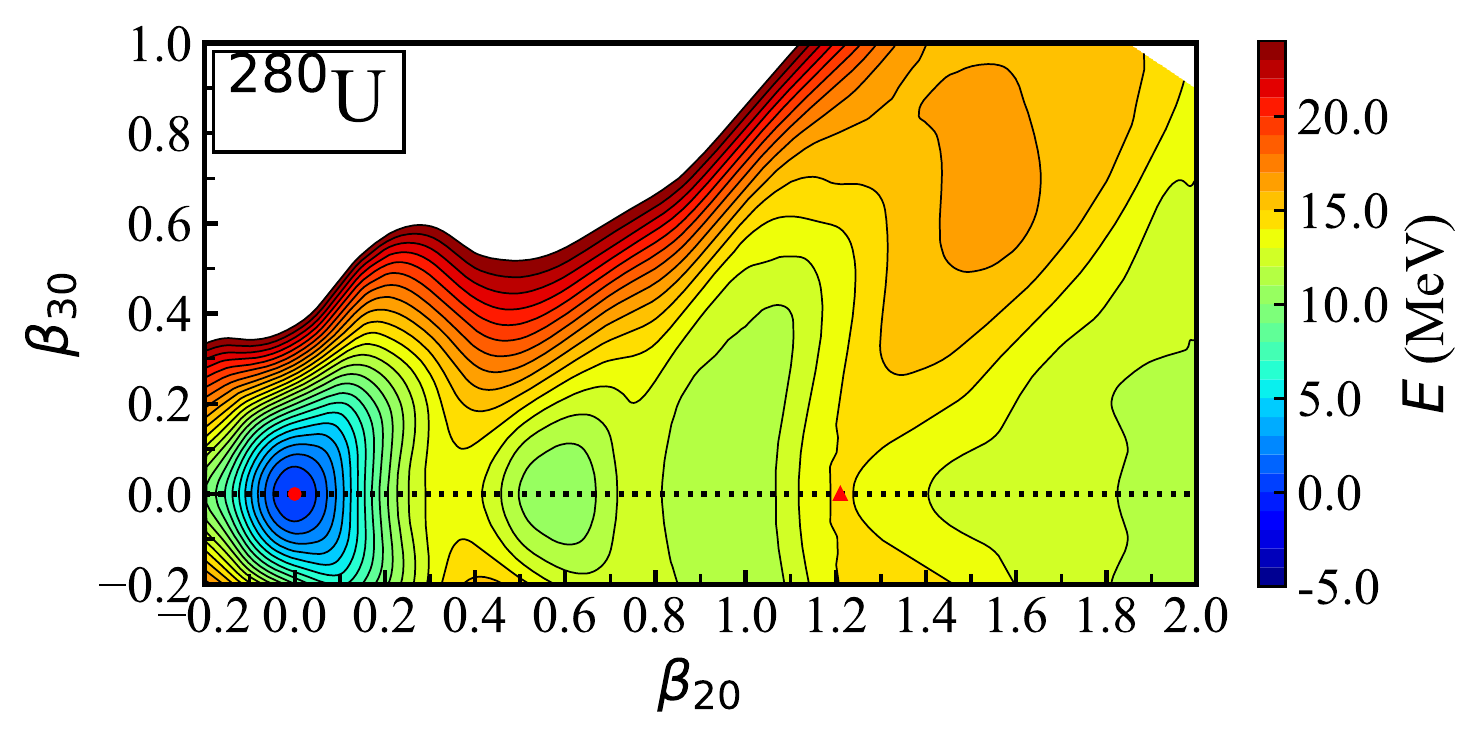}
\includegraphics[width=0.46\textwidth]{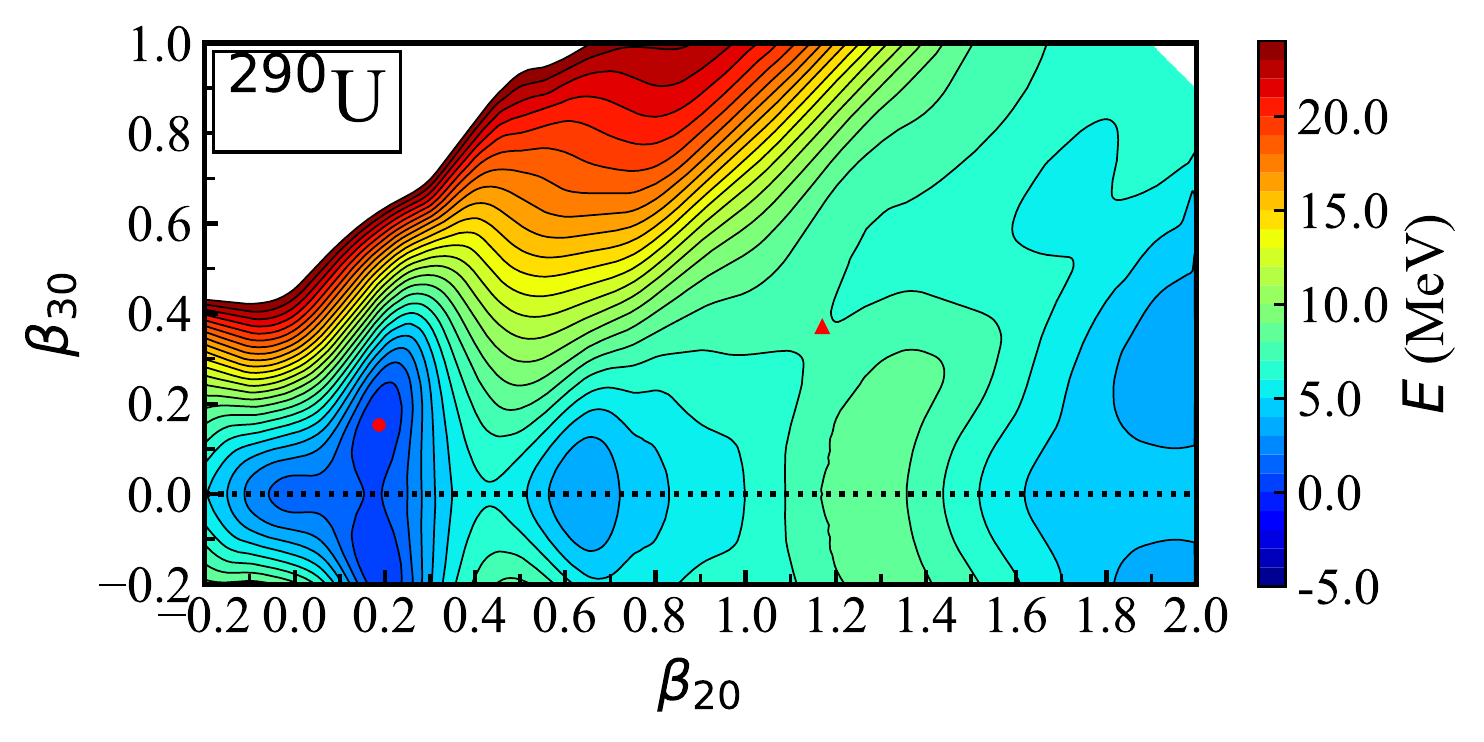}\\
\includegraphics[width=0.46\textwidth]{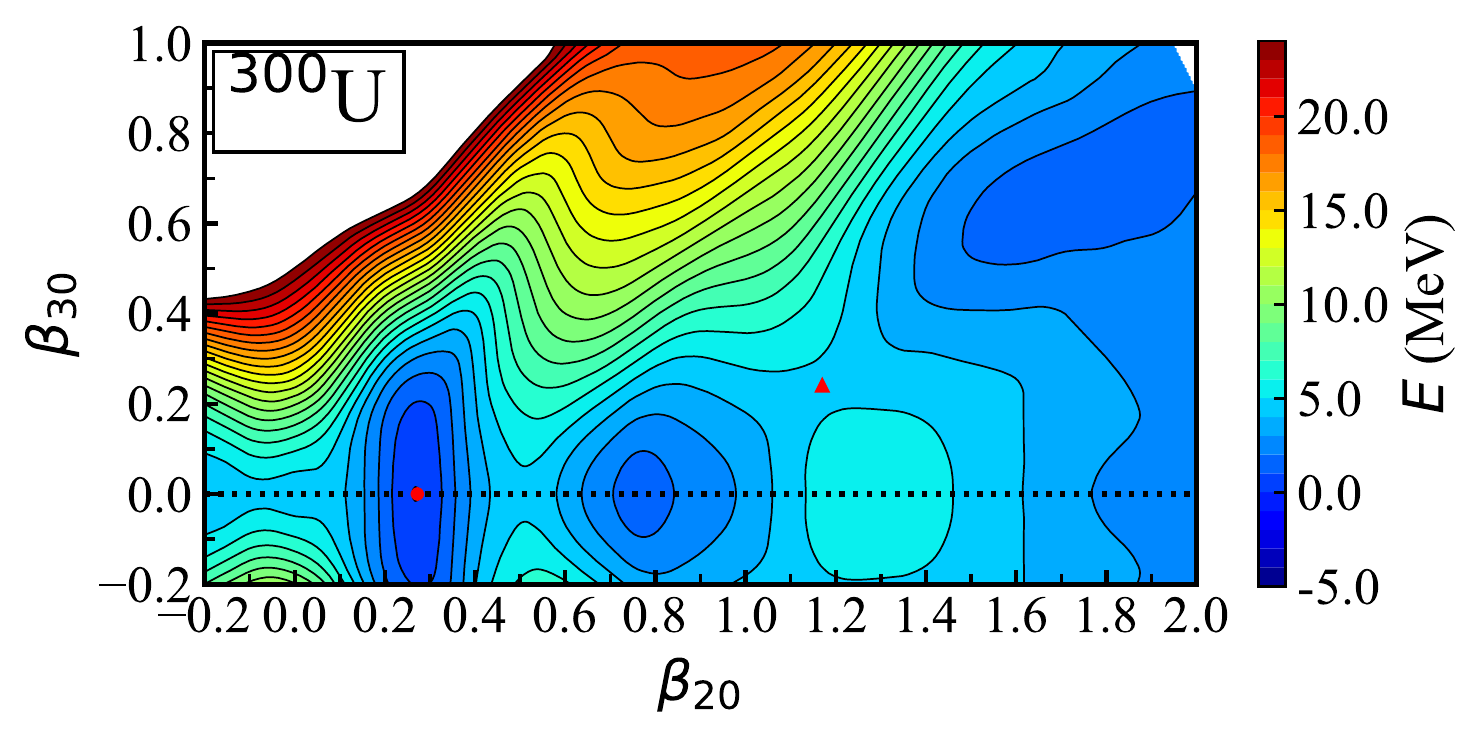}
\includegraphics[width=0.46\textwidth]{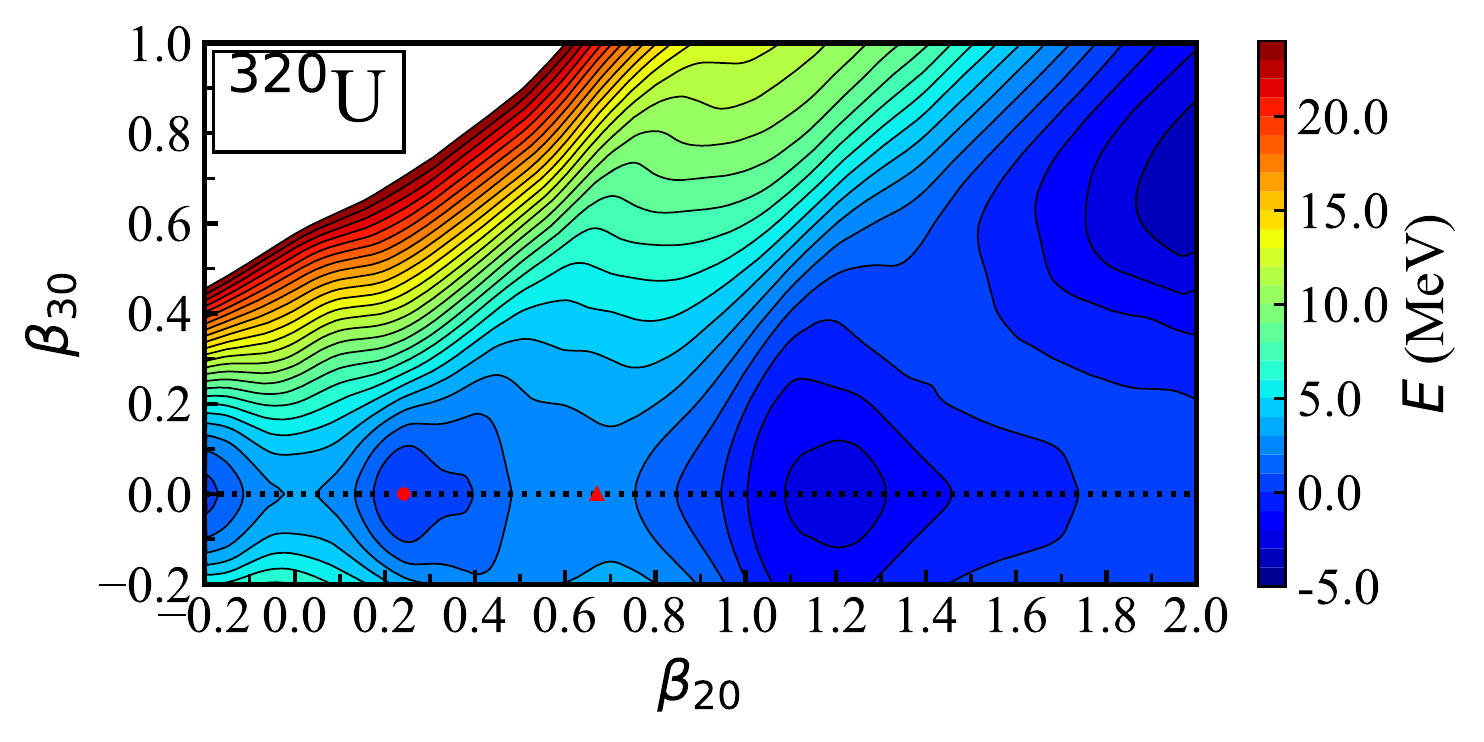}\\
\includegraphics[width=0.46\textwidth]{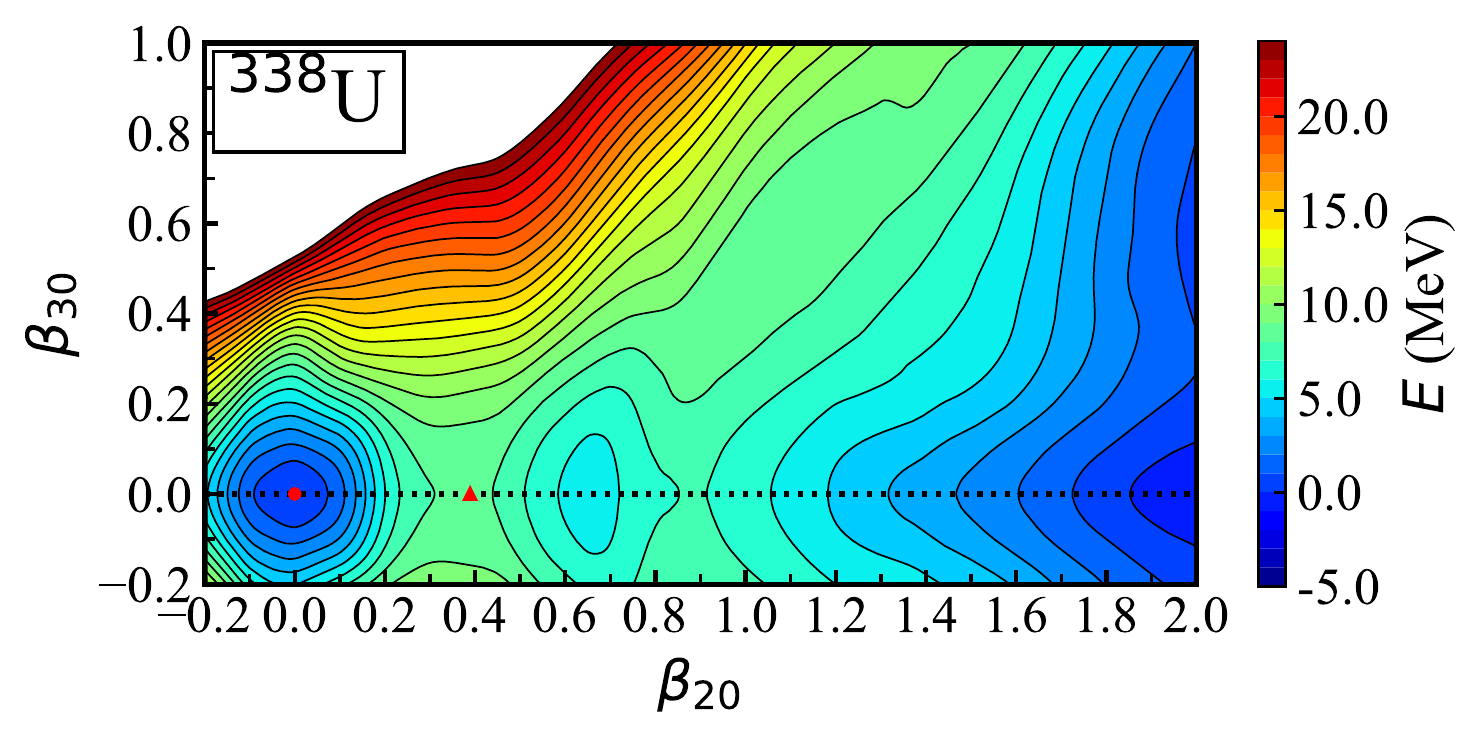}
\includegraphics[width=0.46\textwidth]{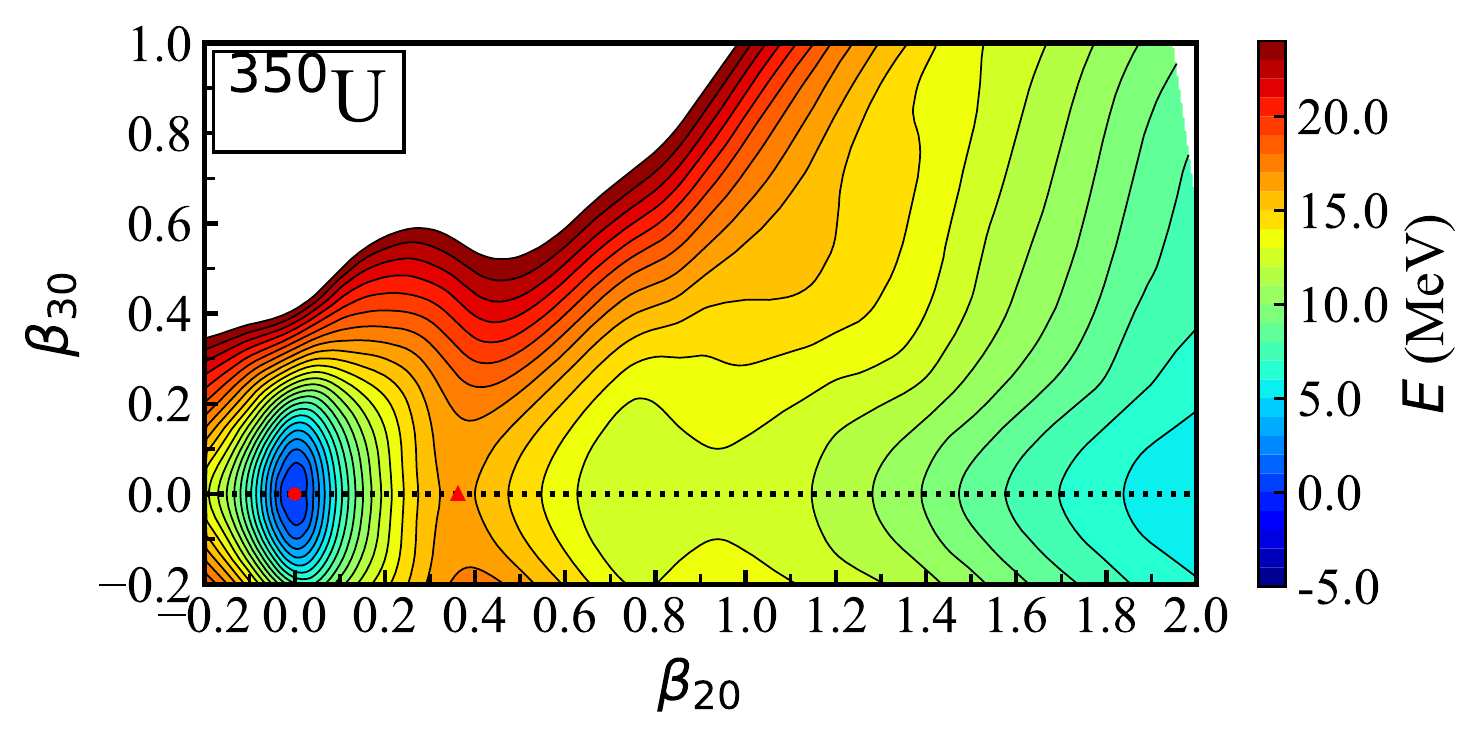}
\end{tabular}
\caption{(Color online)\label{fig:1}
2D PESs of selected U isotopes obtained in the MDC-RMF calculations with the effective interaction PC-PK1 \cite{Zhao2010_PRC82-054319}.
The ground states are indicated by red dots.
The saddle points corresponding to primary barriers are shown by red triangles.
The contour interval is 1.0 MeV.}
\end{figure*}

The PESs of several selected isotopes are shown in Fig.~\ref{fig:1}
where the ground states and the highest saddle points are shown
by red dots and red triangles.
The 2D PESs of U isotopes are mostly smooth and continuous
with obvious minima and saddle points and the static fission pathways
can be easily identified.
First let us take $^{214}$U as an example.
There is a reflection-symmetric (RS) fission pathway and
a reflection-asymmetric (RA) one on the PES,
both starting from a spherical ground state.
There are three saddle points on the RS pathway and two on the RA pathway.
Two shallow third minima can be seen,
meaning that $^{214}$U may fission through both RS and RA pathways.

By examining the PESs of $^{214-250}$U, it is found that the RA fission pathways
are energetically favorable, as can be clearly seen in Fig.~\ref{fig:1}.
For instance, the PES of $^{238}$U is featured by a RS inner barrier,
a low RA outer barrier and a high RS outer barrier which strongly hinder
the nucleus from fissioning through the RS pathway.
Specially, for $^{242-248}$U there are two RA fission pathways starting from
the second minimum; an example, $^{246}$U, is shown in Fig.~\ref{fig:1}.
Similar results were discussed for $^{248}$Cm in Ref.~\cite[Fig.~5]{Lu2012_PRC85-011301R}.
For $^{252-260}$U,
we can find two saddle points on the PESs, corresponding to the typical
double-humped barriers in actinide nuclei.
In these U isotopes, besides the lowest static RS fission path, there is also a RA one, as
seen in the PES of $^{252}$U in Fig.~\ref{fig:1}.
For $^{268-282}$U, the fission barriers are very high.
There are three or four RS barriers on the PESs and
the potential wells between the barriers may correspond to hyperdeformed fission isomers.
The hyperdeformed third or fourth minima will be discussed
in Sec.~\ref{sec:hyperdeformation}.
The RS and RA pathways also appear on the PESs of the isotopes $^{286-334}$U.
From the PESs of $^{336-350}$U, it can be expected these isotopes fission mainly
through the RS pathway.

We can readily locate the ground states of the U isotopes from the 2D PESs.
For most of the isotopes, the ground states are reflection symmetric.
But for $^{226-232}$U and $^{288-294}$U, the $\beta_3$ values are nonzero and
are around 0.2 for $^{228,290}$U as seen in Fig.~\ref{fig:1}.
Detailed ground state properties will be discussed in the next subsection.

\subsection{Ground state properties}

\begin{figure*}
\centering
\begin{tabular}{cc}
\includegraphics[width=0.500\textwidth]{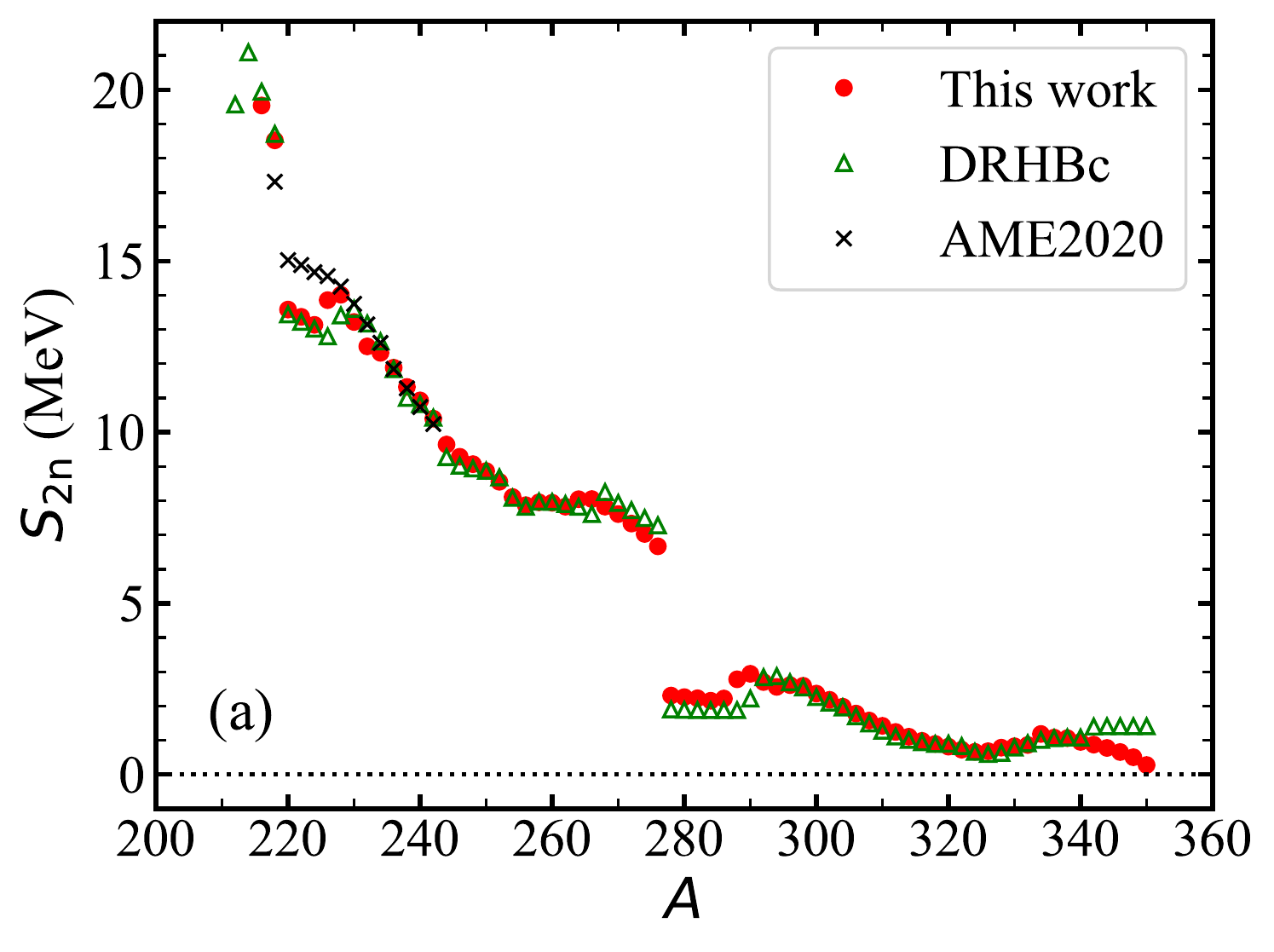}
\includegraphics[width=0.492\textwidth]{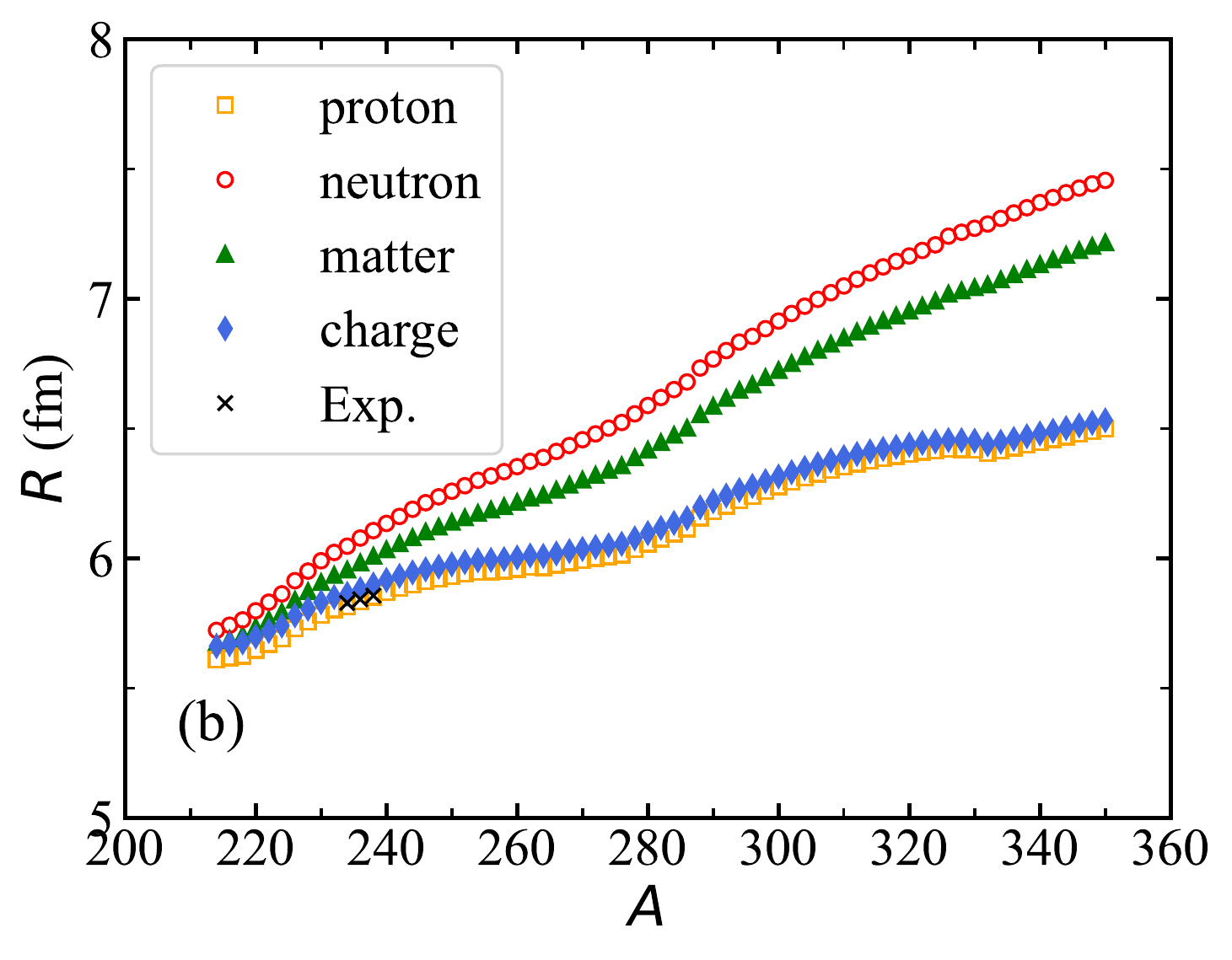}
\end{tabular}
\caption{(Color online) \label{fig:2}
Ground state properties of U isotopes obtained in the MDC-RMF calculations with the effective interaction PC-PK1 \cite{Zhao2010_PRC82-054319}:
(a) Two-neutron separation energies $S_\mathrm{2n}$
compared with available values from the AME2020
\cite{Huang2021_ChinPhysC45-030002,Wang2021_ChinPhysC45-030003} and
the DRHBc mass table \cite{Zhang2022_ADNDT144-101488};
(b) root-mean-square (rms) radii of neutron, proton, matter and charge distributions
as functions of the mass number $A$.
Experimental data for charge radii \cite{Angeli2013_ADNDT99-69} are
shown by the black crosses in (b).}
\end{figure*}

The two-neutron separation energies $S_\mathrm{2n}$ and
root-mean-square (rms) radii of neutron, proton, matter and charge distributions
for U isotopes are shown in Fig.~\ref{fig:2}.

In Fig.~\ref{fig:2}(a), the two-neutron separation energies are compared with
available experimental or evaluated values from the AME2020
\cite{Huang2021_ChinPhysC45-030002,Wang2021_ChinPhysC45-030003} and
the results calculated from the deformed relativistic Hartree–Bogoliubov theory in
continuum (DRHBc) \cite{Zhang2022_ADNDT144-101488}.
Since the reflection symmetry is imposed in the DRHBc theory,
the octupole deformations in the ground states of some U isotopes would certainly
result in differences in the binding and separation energies
between the MDC-RMF model and the DRHBc theory.
The pairing approach and pairing interaction used in these two models are different,
which would also lead to some differences in the ground state properties.
For $^{218}$U,
both the MDC-RMF model and the DRHBc theory overestimate
the experimental $S_\mathrm{2n}$ value and
for $^{220-226}$U, they underestimate the experiments.
In other cases, the MDC-RMF and the DRHBc $S_\mathrm{2n}$ values
are in agreement with the AME2020 whenever available.
As $A$ increases, the $S_\mathrm{2n}$'s from both models share the same trend.
For $342 \le A \le 350$, the MDC-RMF model predicts that
$S_\mathrm{2n}$ decreases monotonically with $A$ increasing and
$^{350}$U is the last bound even-$A$ U isotope.
However, in the DRHBc mass table, $S_\mathrm{2n} = 1.4$ MeV for these five U isotopes
though $^{350}$U is also predicted to be the last bound even-$A$ U isotope.
Such differences concerning $S_\mathrm{2n}$ between these two models
and the different predictions concerning the lightest bound
U isotopes, i.e., $^{212}$U by the DRHBc theory \cite{Zhang2022_ADNDT144-101488}
and $^{214}$U by the MDC-RMF model,
may stem from
that the DRHBc theory can provide a proper description of exotic nuclei
by considering the deformation and continuum effects
\cite{Zhou2010_PRC82-011301R,Li2012_PRC85-024312,Chen2012_PRC85-067301}.

The root-mean-square (rms) radii of neutron, proton, matter and charge distributions
of the ground states are shown in Fig.~\ref{fig:2}(b).
The available experimental values for the charge radii \cite{Angeli2013_ADNDT99-69}
are also included for comparison.
The neutron radii grows faster than proton as the nucleus gets heavier,
leading to thicker neutron skins.
The charge radius $R_{\mathrm{ch}}$ is a significant observable
characterizing the size of a nucleus.
For $^{234}$U, $^{236}$U and $^{238}$U, the experimental values for charge radii are
5.829 fm, 5.843 fm, and 5.857 fm, respectively.
The charge radii calculated with the MDC-RMF model for the three isotopes are
5.864 fm, 5.882 fm and 5.900 fm and are agree with the data within 1\%.
However it is clear that the MDC-RMF model overestimates the data systematically.
Note that such systematic overestimation of the charge radii for
$^{234,236,238}$U ($R_{\mathrm{ch}}$ = 5.863 fm, 5.882 fm and 5.897 fm)
also exists in the DRHBc calculations \cite{Zhang2022_ADNDT144-101488}.
Although an accuracy of 1\% is acceptable for microscopic models
like MDC-RMF and DRHBc, such systematic discrepancy may hint
some insights concerning the possible improvements of these models.

\begin{figure*}[htbp]
\centering
\begin{tabular}{c}
\includegraphics[width=0.58\textwidth]{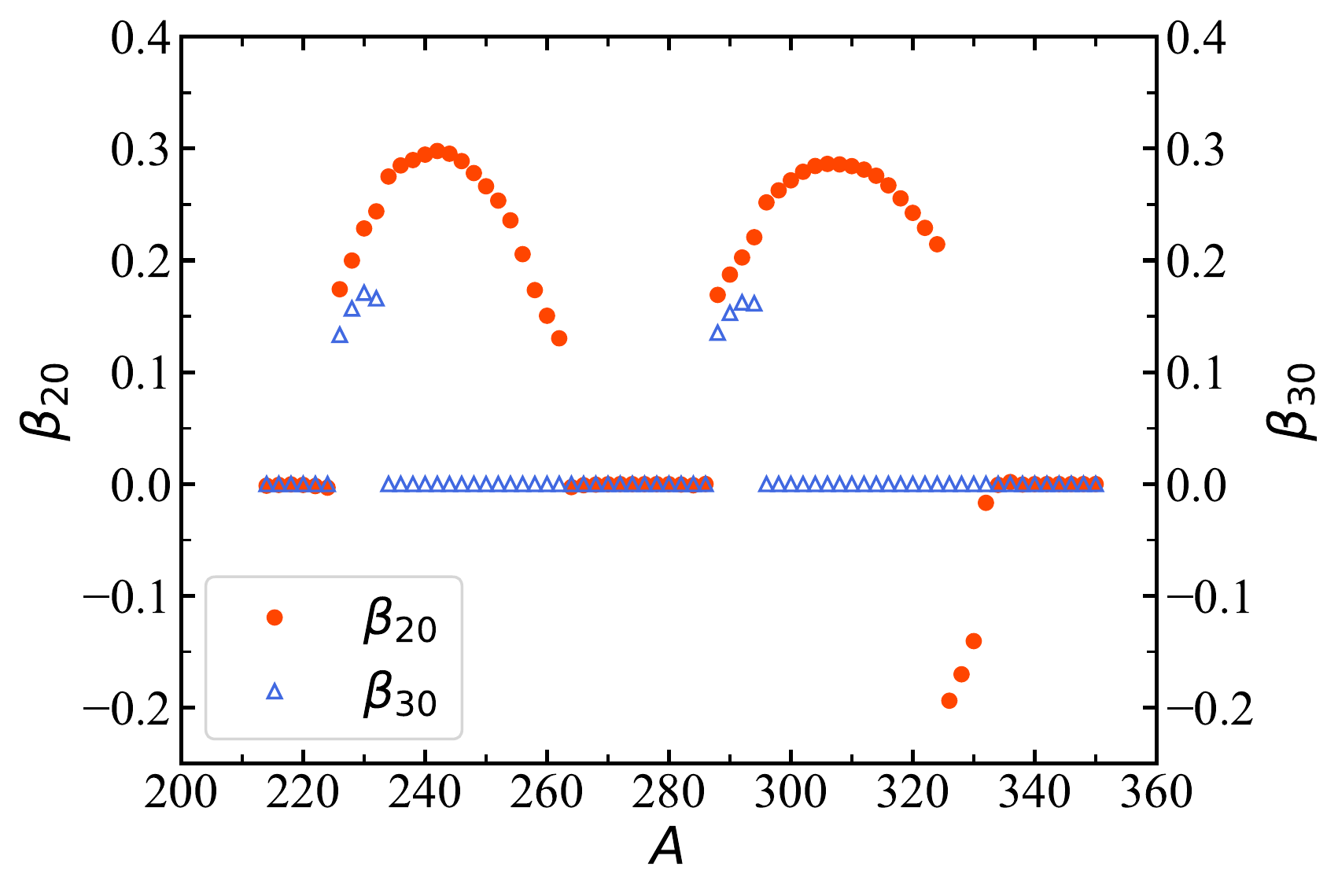}
\end{tabular}
\caption{(Color online)
Quadrupole and octupole deformation parameters of even-$A$ U isotopes as functions of the mass number $A$ obtained in the MDC-RMF calculations with the effective interaction PC-PK1 \cite{Zhao2010_PRC82-054319}.}
\label{fig:3}
\end{figure*}

As we have mentioned in the previous subsection,
quite a large number of U isotopes have non-spherical ground state shapes.
We plot the quadrupole and octupole deformation parameters
($\beta_{20}$ and $\beta_{30}$) of the ground states in Fig.~\ref{fig:3}.
It can be seen that spherical, prolate RA and prolate RS ground state shapes
appear alternately as the mass number increases.
For $^{214-224}$U, $^{264-286}$U and $^{334-350}$U, the ground state shapes are spherical.
For $^{226-262}$U, $\beta_{20}$ increases to a maximum then decreases gradually,
corresponding to various prolate shapes.
In this mass region, $\beta_{20}$ peaks at $^{242}$U with the maximal value 0.30.
Similarly, in the region of $^{288-324}$U, $^{306}$U and $^{308}$U both have
prolate ground state shapes with $\beta_{20}=0.29$.
From $^{326}$U to $^{332}$U, the ground state evolves from
a largely deformed oblate shape to a nearly spherical one,
with $\beta_{20} = -0.19$ and $-0.02$ for $^{326}$U and $^{332}$U, respectively.
From Fig.~\ref{fig:3}, we find that for eight isotopes, i.e., $^{226-232}$U and $^{288-294}$U,
both the quadrupole and octupole deformations appear in their ground states.
These isotopes are featured by pear-shaped ground states.

\begin{figure}
\centering
\begin{tabular}{cc}
\includegraphics[width=0.48\textwidth]{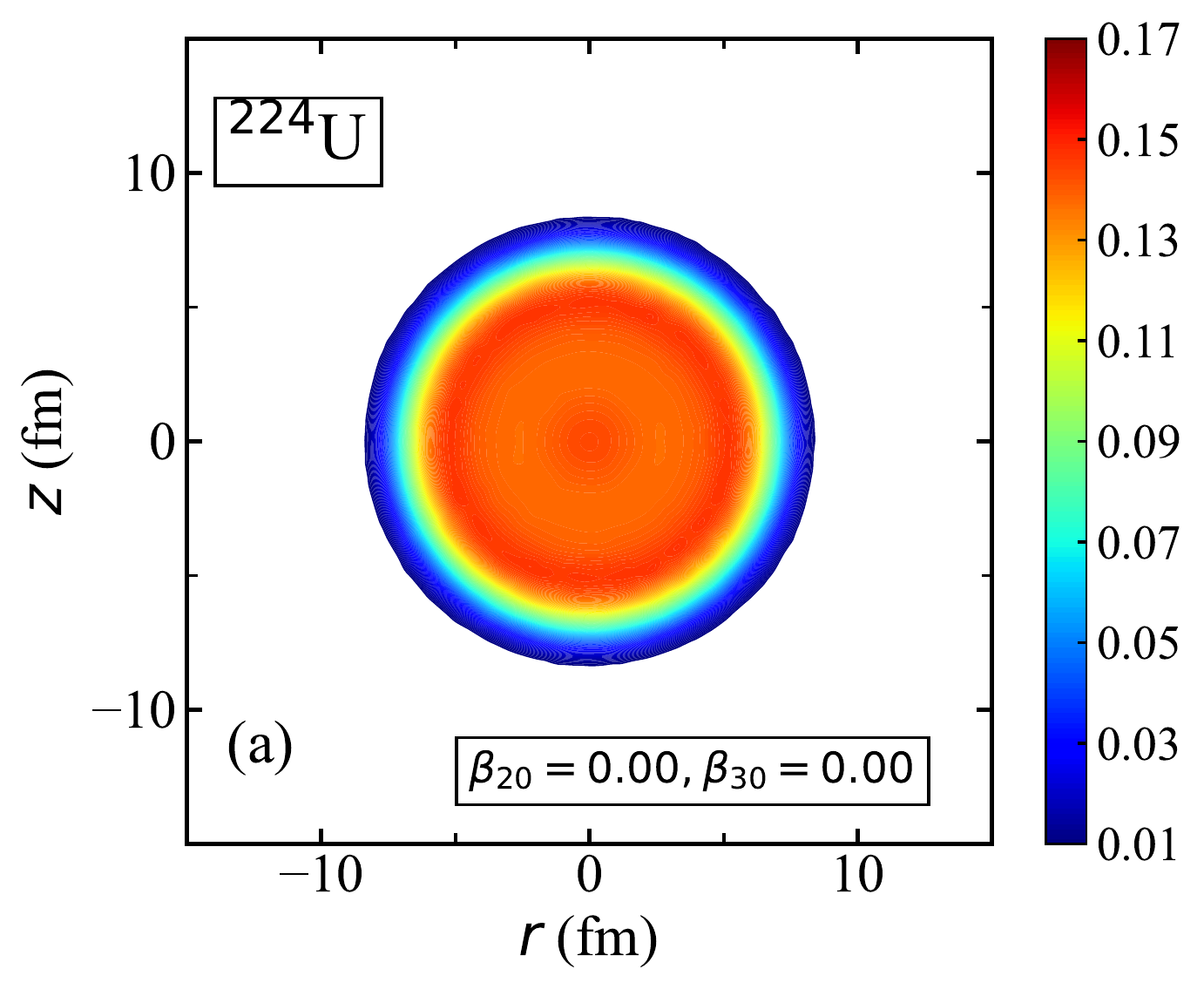}
\includegraphics[width=0.48\textwidth]{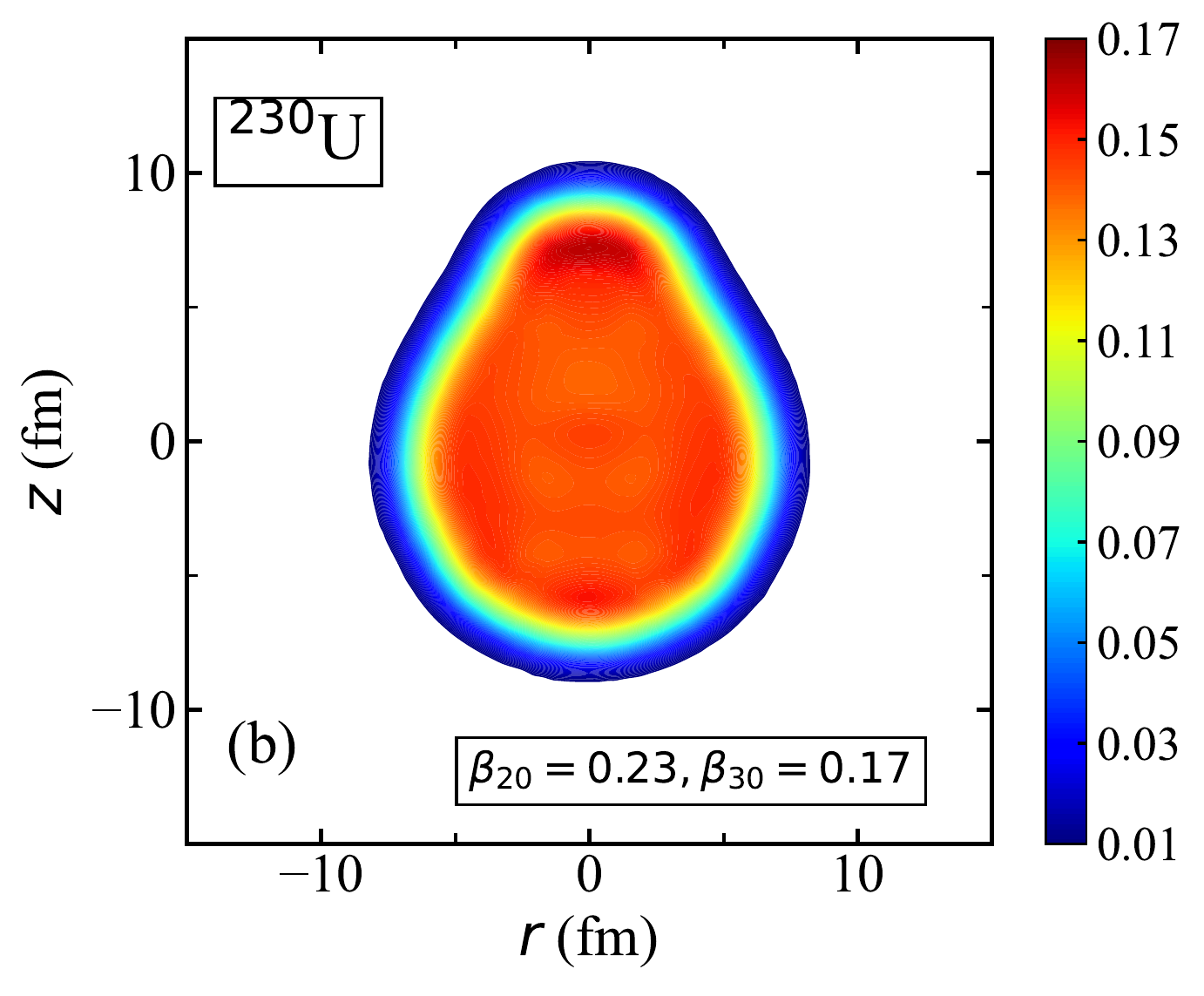}\\
\includegraphics[width=0.48\textwidth]{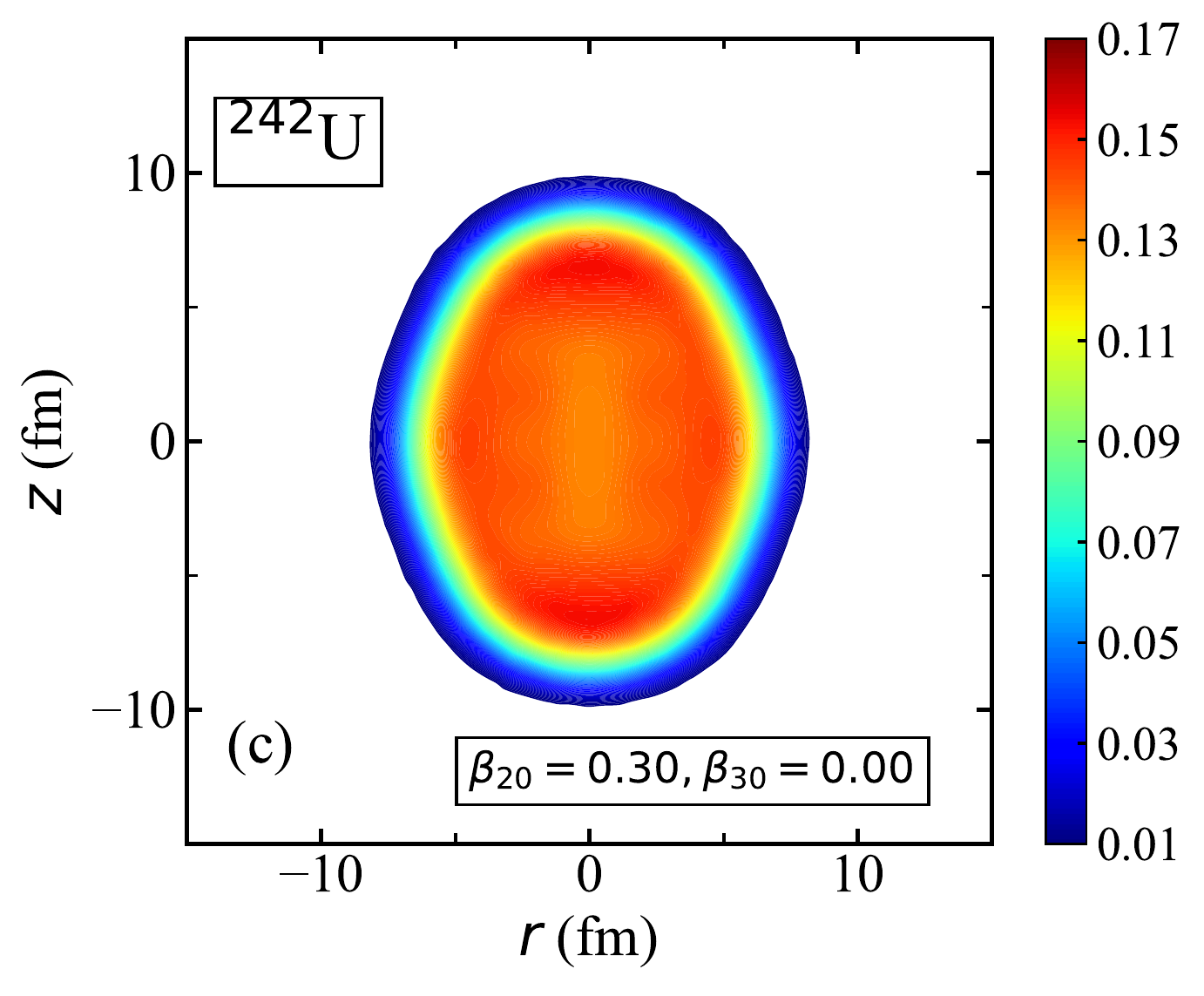}
\includegraphics[width=0.48\textwidth]{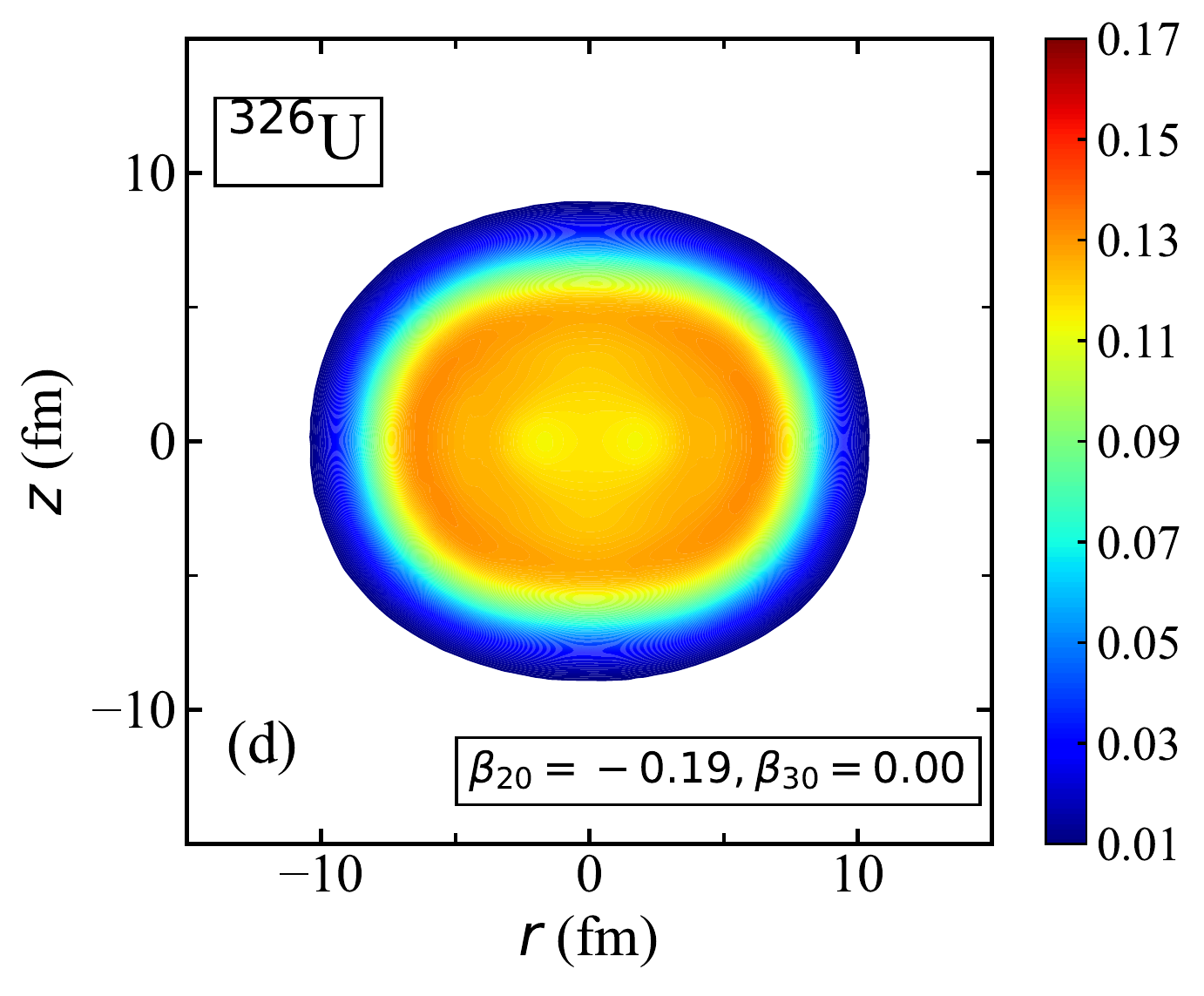}
\end{tabular}
\caption{(Color online)
Density distribution profiles of the ground states of
(a) $^{224}$U, (b) $^{230}$U, (c) $^{242}$U and (d) $^{326}$U obtained in the MDC-RMF calculations with the effective interaction PC-PK1 \cite{Zhao2010_PRC82-054319}.
The $z$-axis is the symmetry axis.
The quadrupole and octupole deformation parameters of the nuclear shapes are shown.}
\label{fig:4}
\end{figure}

In Fig.~\ref{fig:4}, we display density distributions of four typical nuclei
with different kinds of ground state shapes.
The ground states of $^{224}$U, $^{242}$U, and $^{326}$U have a spherical shape,
an axially symmetric prolate shape and an axially symmetric oblate shape, respectively.
The ground state of $^{230}$U has a pear shape with
$\beta_{20} = 0.23$ and $\beta_{30} = 0.17$.

\subsection{Primary barrier heights calculated with octupole and triaxial deformations considered}

\begin{figure*}[htbp]
\centering
\begin{tabular}{c}
\includegraphics[width=0.49\textwidth]{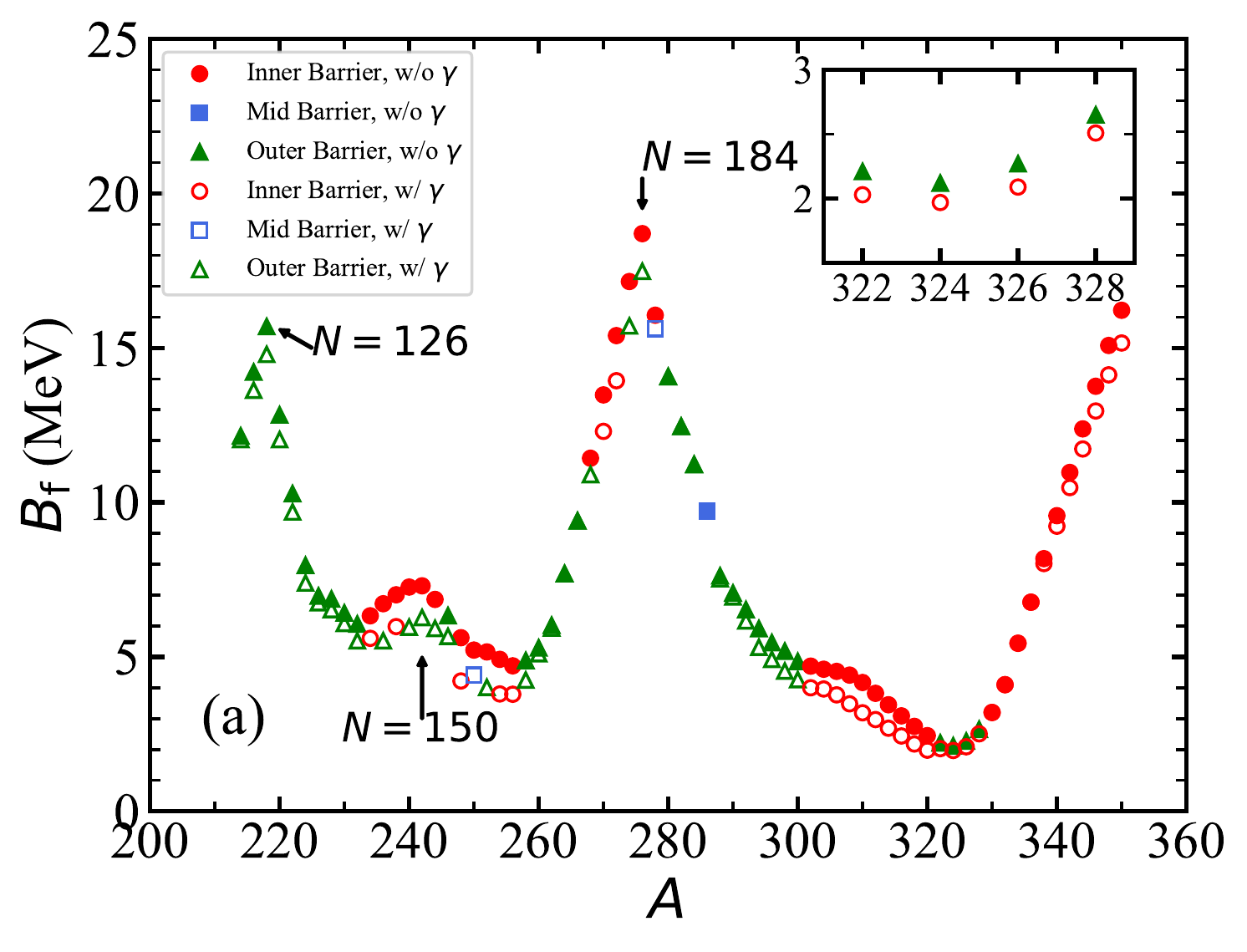}
\includegraphics[width=0.49\textwidth]{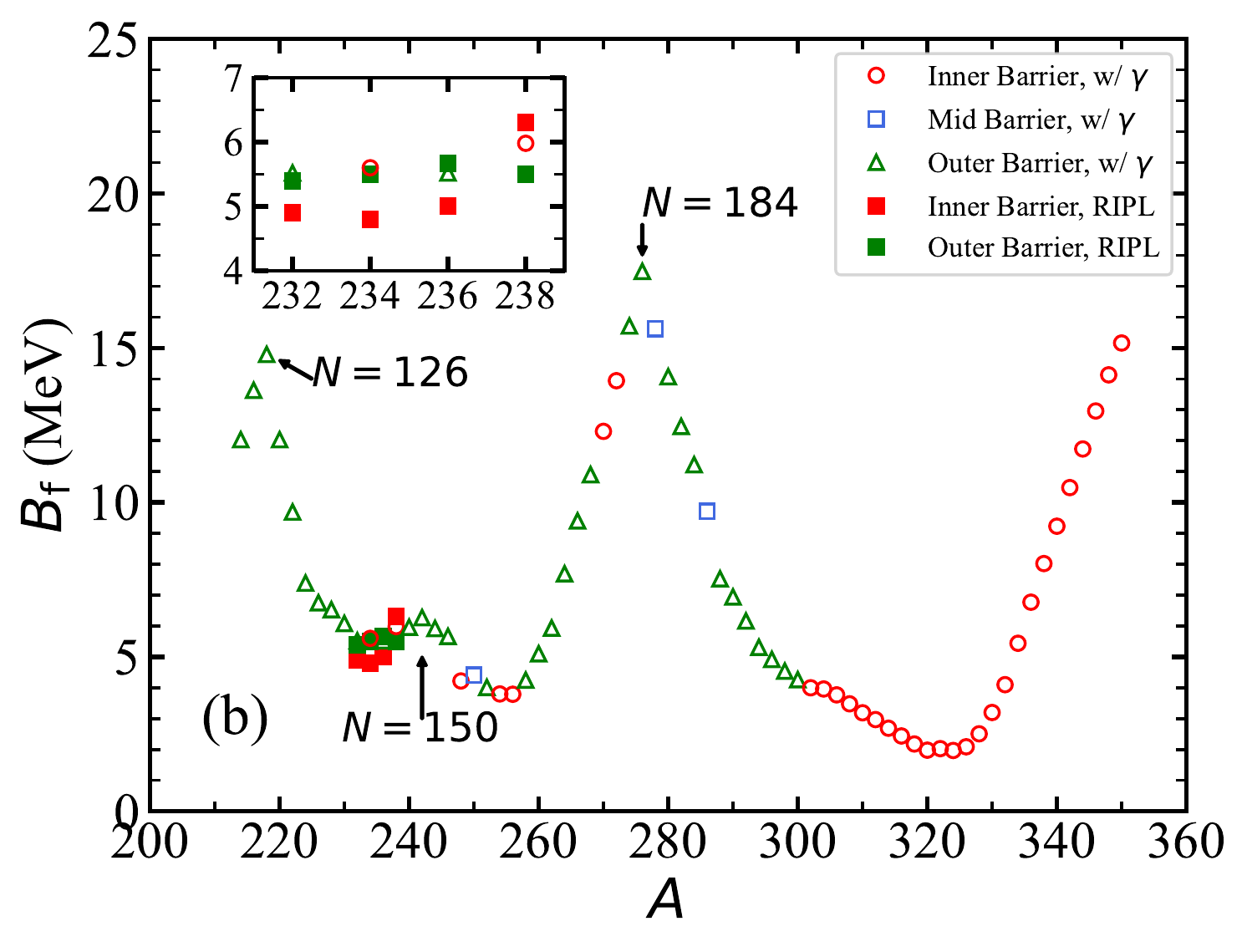}\\
\includegraphics[width=0.49\textwidth]{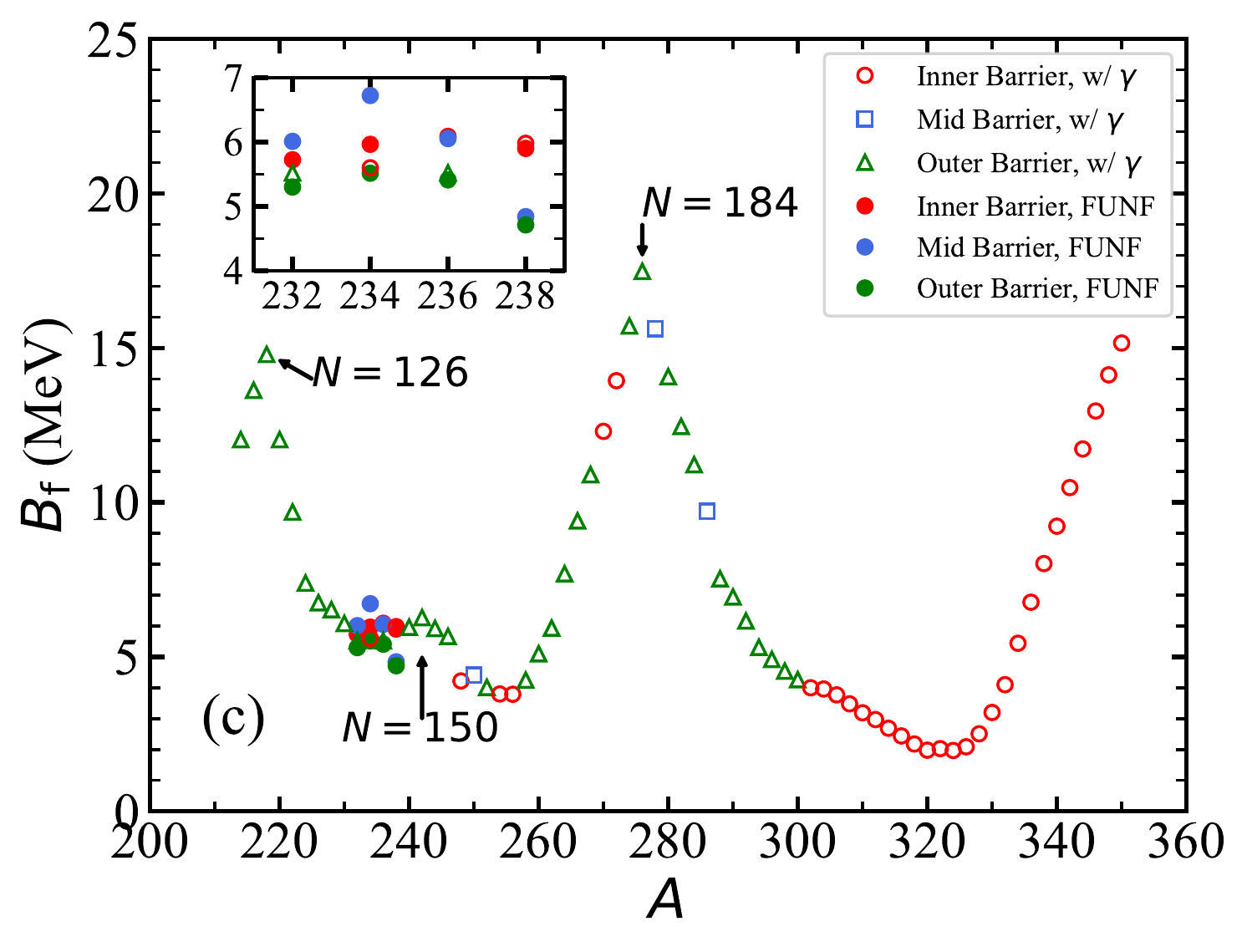}
\includegraphics[width=0.49\textwidth]{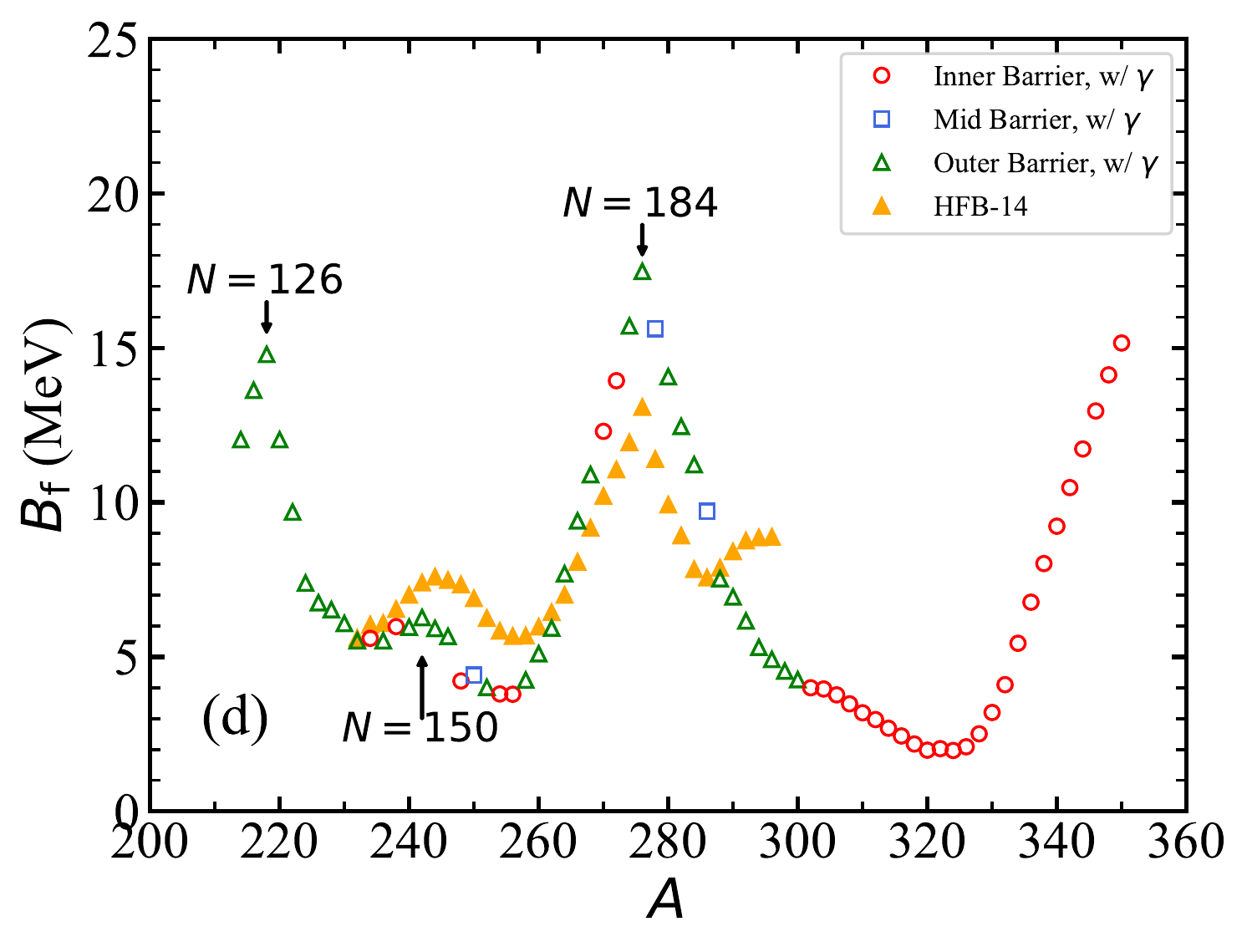}
\end{tabular}
\caption{(Color online)
(a) Primary barrier heights with and without triaxiality obtained in the MDC-RMF calculations with the effective interaction PC-PK1 \cite{Zhao2010_PRC82-054319} and
comparisons of the barrier heights (b) with empirical values from RIPL-3 \cite{Capote2009_NDS110-3107},
(c) with the results given by the FUNF program \cite{Zhang1993_NSE114-55,Zhang2012_FUNF-Manual,Sun2022_PriCom}
and (d) with HFB-14 calculations \cite{Goriely2009_PRC79-024612}.
For the MDC-RMF calculation results, we symbolize the barrier height with
a red circle, blue square or green triangle if the primary barrier of the isotope is
the inner barrier, one of the middle barriers or the outer barrier, respectively.
The barrier heights without and with triaxiality are shown by full symbols
and hollow symbols in (a).
The empirical values of the barrier heights and the FUNF and the HFB-14
results are shown by full symbols in (b), (c) and (d), respectively.}
\label{fig:5}
\end{figure*}

Next we focus on the primary barrier, the highest one, in each even-$A$ U isotope.
The height of the primary barrier is defined as the energy difference
between the highest saddle point and the ground state.
With the calculation results of the 2D PESs we can find all saddle points
for each isotope and determine the primary barrier height.
As seen in the PESs, for many isotopes, the primary barrier in the RA pathway
is obviously lower than the primary barrier in the RS pathway,
which means that the primary barrier height of the nuclei is lowered
when the octupole distortion is allowed.

As we have mentioned, the triaxiality may further lower the height of a barrier
\cite{Moeller1973_IAEA-SM-174-202,Lu2012_PRC85-011301R}.
So in addition to the octupole deformation,
we must take the triaxial deformation into consideration to give
a more accurate description of the primary barrier heights.
Since the MDC-RMF calculations are very time-consuming in
the whole ($\beta_{20}$,$\beta_{22}$,$\beta_{30}$) deformation space,
we break the axial symmetry only in the vicinity of each saddle point
to take the lowering effect of triaxiality on the barriers into account.
With the octupole deformation considered,
the primary barrier heights of U isotopes with and without triaxiality
are shown in Fig.~\ref{fig:5}(a).
For most of the U isotopes, the triaxiality lowers the height of the primary barrier.
The primary barrier may ``shift'' from one barrier to another
when the triaxiality is considered because the lowering effects of
the triaxial distortion for different barriers may be different.
But the general trend of the primary barrier heights versus $A$ is unchanged.
No matter whether we break the axial symmetry or not,
the barrier height peaks at three isotopes with $N$ = 126, 150 and 184,
corresponding to neutron magic numbers and neutron subshells.
For $N=184$, i.e., $^{276}$U, the primary barrier height is as large as 17.48 MeV,
leading to a relatively high stability against fission.

We compare the primary barrier heights with available empirical values taken
from RIPL-3 \cite{Capote2009_NDS110-3107} in Fig.~\ref{fig:5}(b).
The calculated primary barrier heights agree with the RIPL-3 data for
$^{232}$U, $^{236}$U and $^{238}$U;
in both our calculation and RIPL-3,
for $^{232}$U and $^{236}$U the primary barrier is the outer one and
for $^{238}$U the inner one.
The barrier height for $^{234}$U is reproduced by our calculation,
but we predict the inner barrier as the primary one for this isotope.

Based on the FUNF program \cite{Zhang1993_NSE114-55,Zhang2012_FUNF-Manual},
the fission properties of some actinides
have been investigated in detail \cite{Sun2022_PriCom}.
For each isotope, the empirical heights of the inner, middle
and outer barriers are given.
The empirical barrier heights of $^{232-238}$U and our calculation results
are shown together in Fig.~\ref{fig:5}(c).
The heights of outer barriers for $^{232,236}$U and
the inner barrier for $^{238}$U given by the MDC-RMF model are
in agreement with the FUNF results.
The inner barrier for $^{234}$U from our calculation is about 0.5 MeV
lower than that from the FUNF program.
However, in most cases the primary barriers predicted by the MDC-RMF model
are different from the FUNF results.
Only for $^{238}$U, both the MDC-RMF model and the FUNF program show that
the inner barrier is the primary one.
The FUNF calculations predict that the middle barrier is the primary one
for $^{232}$U and $^{234}$U.
For $^{236}$U, the FUNF program shows that the inner barrier is the
highest among the investigated barriers.
It is worth noting that the FUNF program predicts the appearance of
a middle barrier with a considerable height for $^{232-236}$U.
Such results for the middle barriers are different from those given by
the MDC-RMF calculations.

Based on other microscopic models, the barrier heights of the U isotopic chain
are also investigated.
In Fig.~\ref{fig:5}(d), the comparison of the MDC-RMF results with fission
barrier heights for $^{232-296}$U given by HFB-14 calculations
\cite{Goriely2009_PRC79-024612} is shown.
For $^{232-286}$U, the barrier heights from the MDC-RMF model and
the HFB-14 method show the same trend as the mass number increases.
For $^{232-262}$U and $^{288-296}$U, the HFB-14 method gives higher barriers
in comparison with the MDC-RMF model; while
in $^{264-286}$U, our calculation results predict higher barriers.
A drastic difference between the two model calculation results
can be seen for the mass region $286 \le A \le 296$:
The HFB-14 barrier heights increase with $A$ while
the MDC-RMF values decrease.
The reason behind such a different behavior is yet not clear to us
and should be studied in the future.

\subsection{\label{sec:hyperdeformation}
Hyperdeformed third and fourth minima on the PESs}

\begin{figure*}[htbp]
\centering
\begin{tabular}{c}
\includegraphics[width=0.58\textwidth]{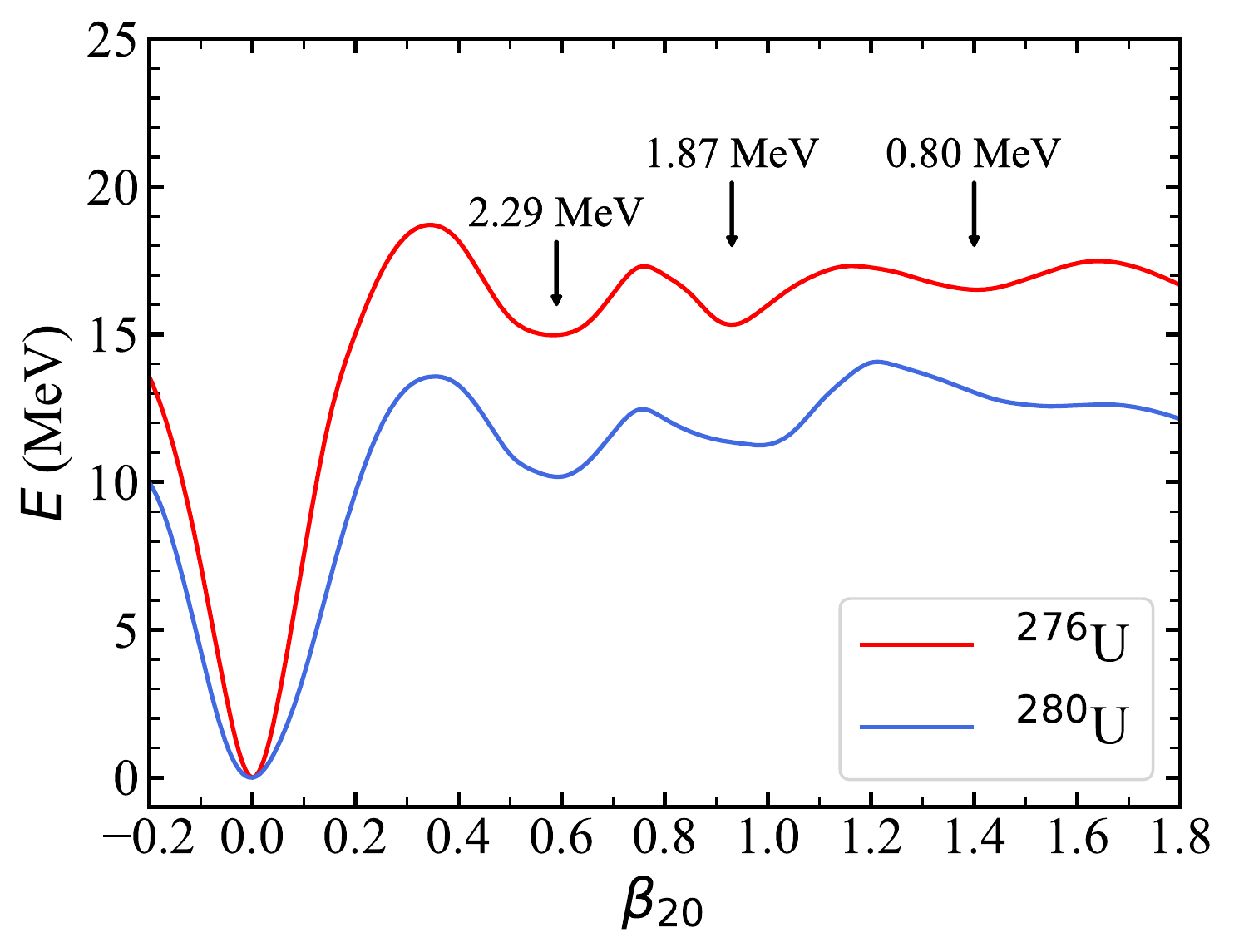}
\end{tabular}
\caption{(Color online)
One-dimensional potential energy curves of $^{276}$U and $^{280}$U obtained in the MDC-RMF calculations with the effective interaction PC-PK1 \cite{Zhao2010_PRC82-054319}.
For $^{276}$U, the positions and depths (the energy difference between the minimum and the lower one of the two saddle points around it) of the second, third and fourth
potential wells are indicated.
The curves represent fission pathways with the axial symmetry and the reflection symmetry imposed.}
\label{fig:6}
\end{figure*}

By investigating the potential energy curves or surfaces of some actinides,
the appearance of third minima corresponding to hyperdeformed nuclear shapes
has been reported and many experiment efforts have also been made
\cite{Cwiok1994_PLB322-304,Rutz1995_NPA590-680,
Krasznahorkay1998_APHA7-35,Marinov2001_IJMPE10-209,Thirolf2001_APHA13-111,
Thirolf2002_PPNP49-325,Krasznahorkay2003_APHA18-323,
Delaroche2006_NPA771-103,Sin2006_PRC74-014608,Kowal2012_PRC85-061302R,
Ichikawa2013_PRC87-054326,Jachimowicz2013_PRC87-044308,
McDonnell2013_PRC87-054327,Zhao2015_PRC91-014321}.
From our MDC-RMF calculation results for the even-$A$ U isotopes,
we can find obvious RS third and even fourth minima in PESs of $^{268-282}$U.
As examples, we show the potential energy curves $E(\beta_{20})$
at $\beta_{30}=0$ for $^{276}$U and $^{280}$U in Fig.~\ref{fig:6}.
$^{276}$U, the isotope with neutron magic number $N=184$,
has the highest fission barriers among the U isotopes.
The height of the inner barrier is 18.8 MeV and the value is lowered
by 1.64 MeV when the triaxial deformations are considered.
The depths (the energy difference between the minimum and the lower one of the two saddle points around it) of the second, third and fourth potential wells
for $^{276}$U are 2.29 MeV, 1.87 MeV and 0.80 MeV.
The positions and depths of these wells are indicated in Fig.~\ref{fig:6}.

The depths of the third well of $^{226,228,230,232}$Th and $^{232,234,238}$U
have been studied by using the MDC-RMF model with functionals DD-ME2
and PC-PK1 \cite{Zhao2015_PRC91-014321}.
With PC-PK1, the third well only appears in $^{226,228,230}$Th and
the depths are 1.29 MeV, 0.78 MeV and 0.44 MeV, respectively.
By investigating the deformed single-particle levels of proton and neutron,
the appearance of the third minimum has been attributed to a proton shell gap
at large deformations
which stems from several pairs of single-proton states in the vicinity
of the Fermi surface \cite{Zhao2015_PRC91-014321}.
In comparison with the results given in Ref.~ \citenum{Zhao2015_PRC91-014321},
the third well of $^{276}$U with a depth of 1.87 MeV is deeper than
those of the previously investigated Th isotopes.
The fourth well of $^{276}$U is even deeper than the third wells of $^{228}$Th and $^{230}$Th.
On the potential energy curve of $^{280}$U in Fig.~\ref{fig:6},
the second and third minima can be identified, but the fourth well
no longer exists due to the absence of a fourth saddle point.

\begin{figure*}[htbp]
\centering
\begin{tabular}{cc}
\includegraphics[width=0.48\textwidth]{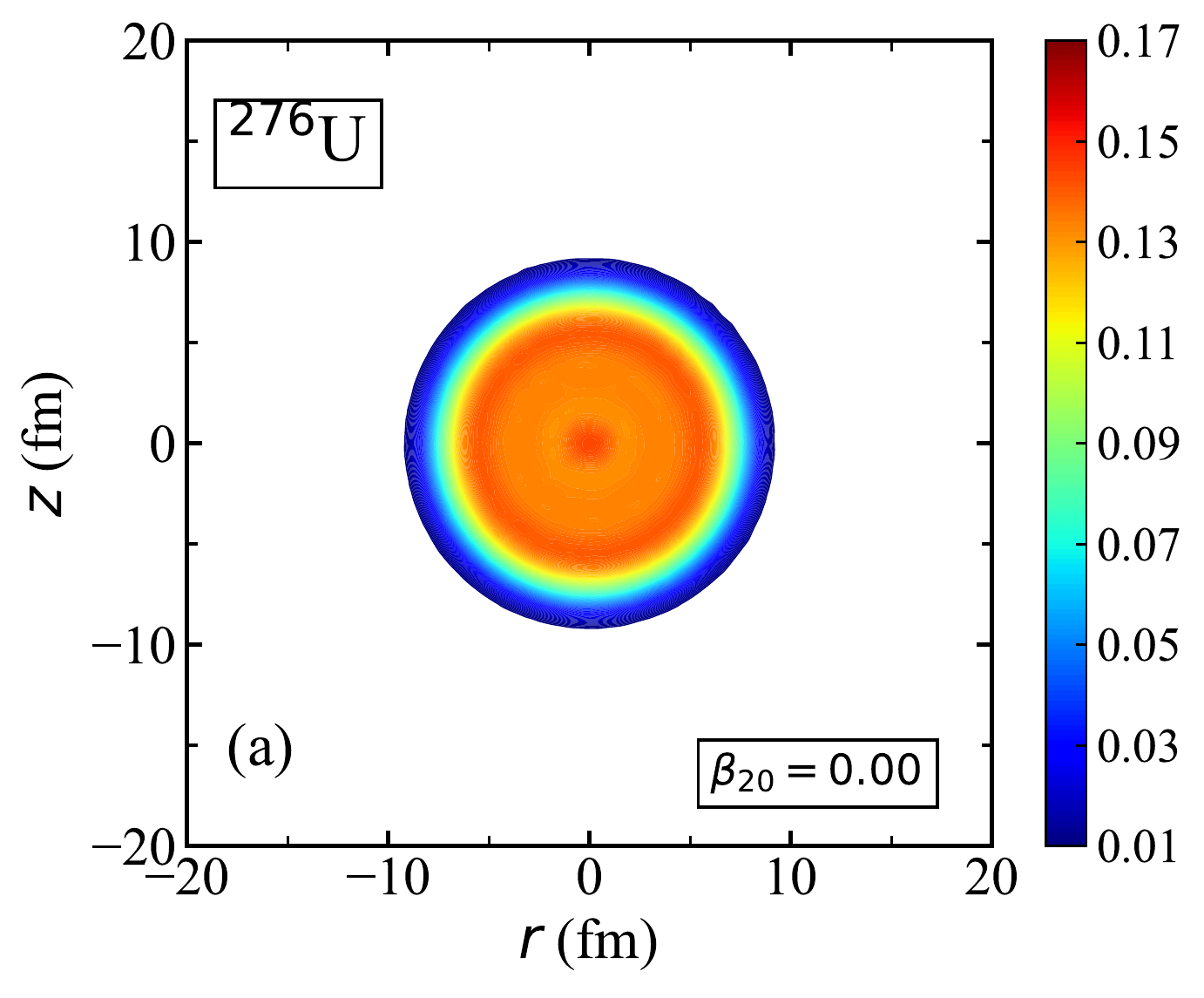}
\includegraphics[width=0.48\textwidth]{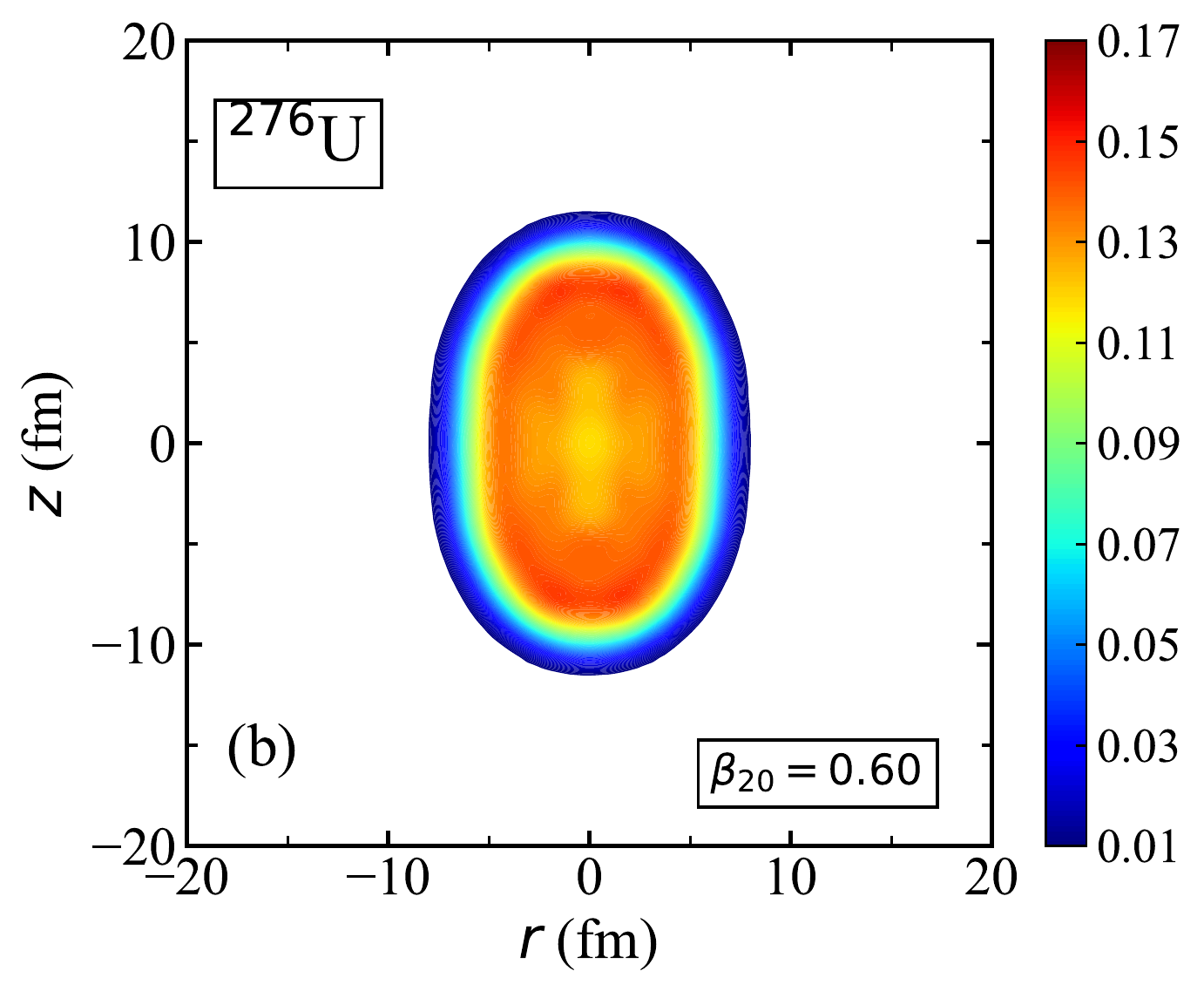}\\
\includegraphics[width=0.48\textwidth]{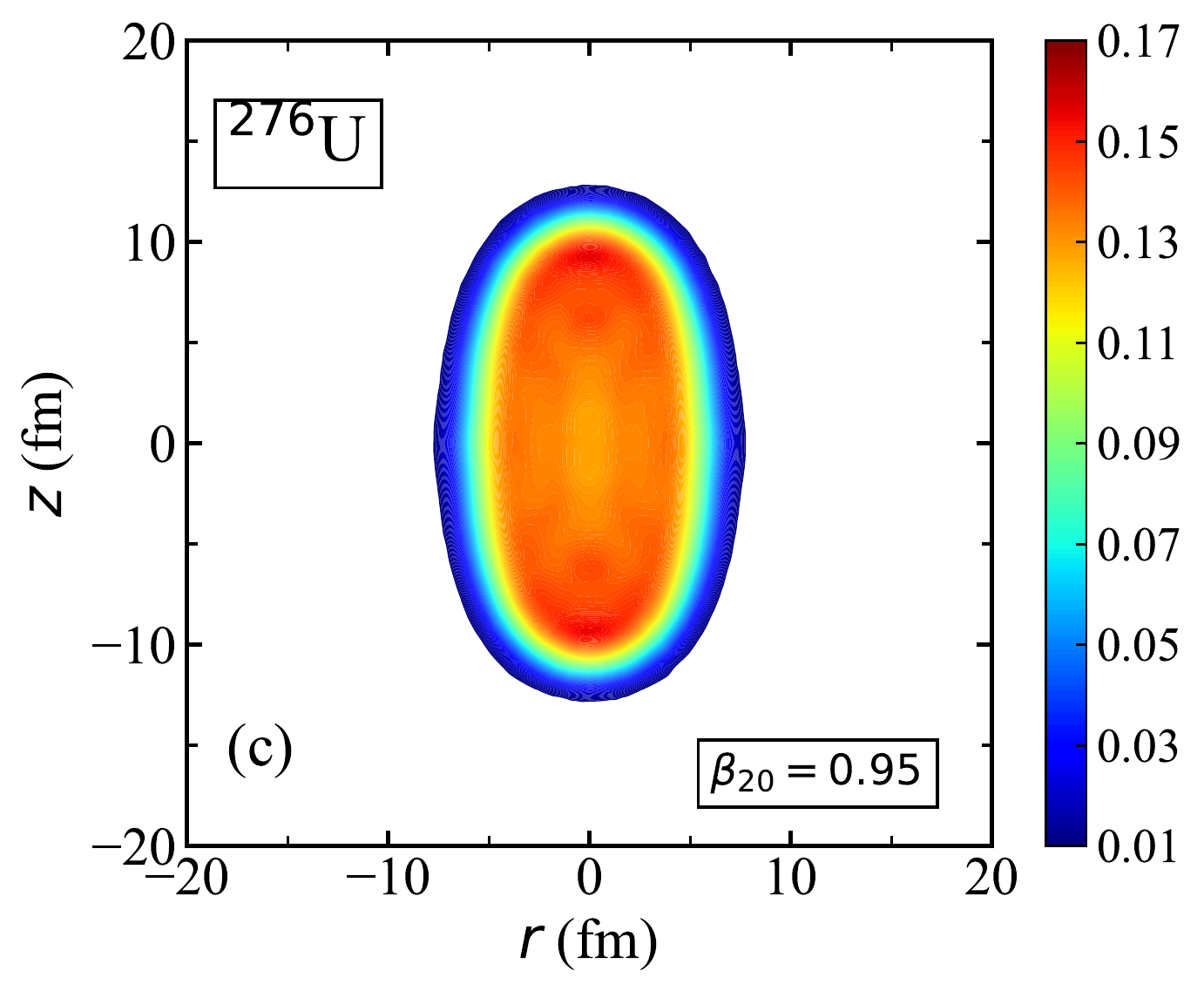}
\includegraphics[width=0.48\textwidth]{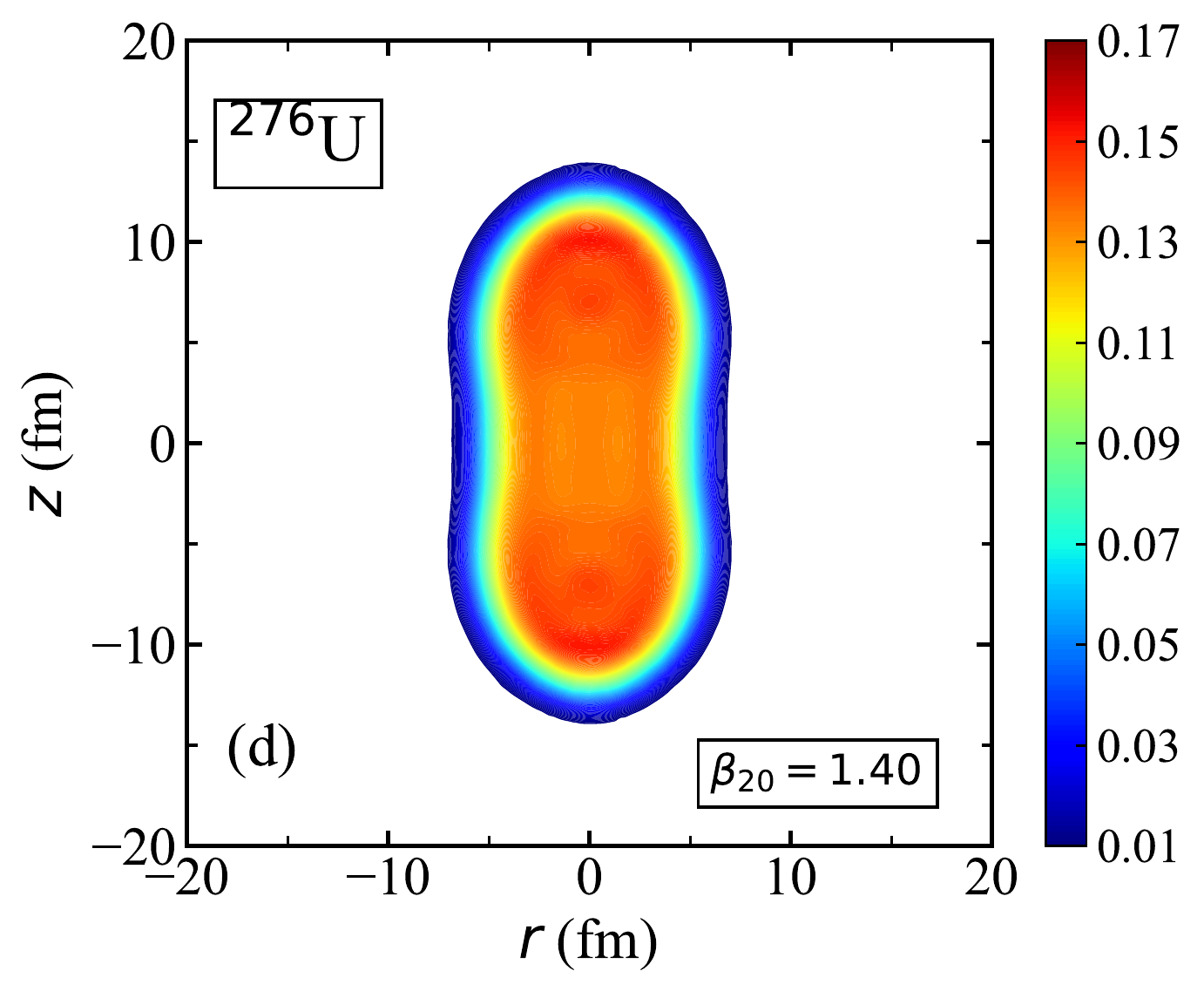}
\end{tabular}
\caption{(Color online)
Density distribution profiles of $^{276}$U at (a) the ground state and (b) the second,
(c) third and (d) fourth minima obtained in the MDC-RMF calculations with the effective interaction PC-PK1 \cite{Zhao2010_PRC82-054319}.
The nucleus is in axial-symmetric shapes at these minima and the quadrupole deformation parameters are shown.}
\label{fig:7}
\end{figure*}

The density distribution profiles of $^{276}$U at the ground state and
the second, third and fourth minima are shown in Fig.~\ref{fig:7}.
The ground state of $^{276}$U is spherical.
At the second minimum, the nucleus is in a prolate shape.
The nucleus further elongates when it is in the third well.
At the fourth minimum, the neck appears and the nucleus shows a tendency to split up.
With $\beta_{20}=0.95$ and $\beta_{20}=1.40$ at the third and fourth minima,
$^{276}$U is highly deformed, corresponding to hyperdeformed fission isomers.

\section{Summary}\label{sec4}

In summary, we have systematically studied the properties of even-$A$ U isotopes
by using the MDC-RMF model.
The PESs of $^{214}$U to $^{350}$U in the ($\beta_{20}$,$\beta_{30}$) plane are
obtained and examined.
Most of the PESs are characterized by a double-humped barrier structure.
But for some isotopes, three or four saddle points appear.
Detailed results concerning the ground state properties,
including two-neutron separation energies,
root-mean-square radii for the neutron, proton, matter and charge distributions and
quadrupole and octupole deformations, are presented.
It can be seen from the calculation results that many of these U isotopes
have non-spherical ground states and several of them,
e.g., $^{230}$U, have a pear-shaped ground state
with large $\beta_{20}$ and $\beta_{30}$.

By considering both the triaxiality and the reflection asymmetry
in the vicinities of the saddle points,
the primary barriers of these U isotopes have been investigated in detail.
For many of them, the octupole deformations lower the primary barrier.
The triaxiality may further lower the barriers and in some cases the primary barrier
may shift from one barrier to another because the lowering effects of
the triaxial distortion for the inner, middle and outer barriers may be different.
The primary barrier heights with octupole and triaxial deformations
can reproduce available empirical values.
The hyperdeformed third and fourth minima on the PESs are discussed.
Taking $^{276}$U as an example, the depths of the third and the fourth potential
well are 1.87 MeV and 0.80 MeV respectively.

\section*{Acknowledgements}

Helpful discussions with
Bao-Ge Deng,
Bing-Nan Lu,
Yu-Ting Rong,
Xiang-Xiang Sun,
Zhong-Hao Tu,
Kun Wang,
Xiao-Qian Wang
and
Zhen-Hua Zhang
are gratefully acknowledged.
We thank Xiao-Jun Sun for providing us the FUNF results prior publication.
This work has been partly supported by
the Strategic Priority Research Program of Chinese Academy of Sciences (Grants No. XDB34010000 and No. XDPB15),
the National Key R\&D Program of China (Grant No. 2018YFA0404402),
the National Natural Science Foundation of China (Grants
No. 11525524, No. 12070131001, No. 12047503, and No. 11961141004),
the Inter-Governmental S\&T Cooperation Project between China and Croatia
and
the IAEA Coordinated Research Project ``F41033''.
The results described in this paper are obtained on
the High-performance Computing Cluster of ITP-CAS and
the ScGrid of the Supercomputing Center, Computer Network Information Center of Chinese Academy of Sciences.


\begin{thebibliography}{100}

\bibitem{Bohr1998_Nucl_Structure_1}
A.~Bohr and B.~R. Mottelson, {\em Nuclear {{Structure}}} ({World Scientific},
  {Singapore}, 1998).

\bibitem{Bohr1998_Nucl_Structure_2}
A.~Bohr and B.~R. Mottelson, {\em Nuclear {{Structure}}} ({World Scientific},
  {Singapore}, 1998).

\bibitem{Ring1980}
P.~Ring and P.~Schuck, {\em The Nuclear Many Body Problem}, Texts and
  Monographs in Physics ({Springer-Verlag}, {New York, Heidelberg, Berlin},
  1980).

\bibitem{Bohr1953_PR089-316}
A.~Bohr and B.~R. Mottelson, {\em Phys. Rev.} {\bf 89}  (1953) 316.

\bibitem{Bohr1953_PR090-717}
A.~Bohr and B.~R. Mottelson, {\em Phys. Rev.} {\bf 90}  (1953) 717.

\bibitem{Alder1956_RMP28-432}
K.~Alder, A.~Bohr, T.~Huus, B.~Mottelson and A.~Winther, {\em Rev. Mod. Phys.}
  {\bf 28}  (1956) 432.

\bibitem{Alder1958_RMP30-353}
K.~Alder, A.~Bohr, T.~Huus, B.~Mottelson and A.~Winther, {\em Rev. Mod. Phys.}
  {\bf 30}  (1958) 353.

\bibitem{Bohr1957_NP4-529}
A.~Bohr and B.~R. Mottelson, {\em Nucl. Phys.} {\bf 4}  (1957) 529.

\bibitem{Bohr1958_NP9-687}
A.~Bohr and B.~R. Mottelson, {\em Nucl. Phys.} {\bf 9}  (1958) 687.

\bibitem{Frauendorf1997_NPA617-131}
S.~Frauendorf and {Jie Meng}, {\em Nucl. Phys. A} {\bf 617}  (1997) 131.

\bibitem{Meng2006_PRC73-037303}
J.~Meng, J.~Peng, S.~Q. Zhang and S.-G. Zhou, {\em Phys. Rev. C} {\bf 73}
  (2006)   037303.

\bibitem{Zhao2015_PRC91-014321}
J.~Zhao, B.-N. Lu, D.~Vretenar, E.-G. Zhao and S.-G. Zhou, {\em Phys. Rev. C}
  {\bf 91}  (2015)   014321.

\bibitem{Zhao2015_PRL115-022501}
P.~W. Zhao, N.~Itagaki and J.~Meng, {\em Phys. Rev. Lett.} {\bf 115}  (2015)
  022501.

\bibitem{Dudek2006_PRL97-072501}
J.~Dudek, D.~Curien, N.~Dubray, J.~Dobaczewski, V.~Pangon, P.~Olbratowski and
  N.~Schunck, {\em Phys. Rev. Lett.} {\bf 97}  (2006)   072501.

\bibitem{Chen2010_NPA834-378c}
Y.~Chen and Z.-C. Gao, {\em Nucl. Phys. A} {\bf 834}  (2010) 378c.

\bibitem{Zhao2017_PRC95-014320}
J.~Zhao, B.-N. Lu, E.-G. Zhao and S.-G. Zhou, {\em Phys. Rev. C} {\bf 95}
  (2017)   014320.

\bibitem{Wang2022_CTP74-015303}
K.~Wang and B.-N. Lu, {\em Commun. Theor. Phys.} {\bf 74}  (2022)   015303.

\bibitem{Wong1973_APNY77-279}
C.~Wong, {\em Annals Phys.} {\bf 77}  (1973) 279.

\bibitem{Cao2019_PRC99-014606}
X.~G. Cao, E.~J. Kim, K.~Schmidt, K.~Hagel, M.~Barbui, J.~Gauthier,
  S.~Wuenschel, G.~Giuliani, M.~R.~D. Rodriguez, S.~Kowalski, H.~Zheng,
  M.~Huang, A.~Bonasera, R.~Wada, N.~Blando, G.~Q. Zhang, C.~Y. Wong,
  A.~Staszczak, Z.~X. Ren, Y.~K. Wang, S.~Q. Zhang, J.~Meng and J.~B. Natowitz,
  {\em Phys. Rev. C} {\bf 99}  (2019)   014606.

\bibitem{Bohr1939_PR056-426}
N.~Bohr and J.~A. Wheeler, {\em Phys. Rev.} {\bf 56}  (1939) 426.

\bibitem{Strutinsky1967_NPA95-420}
V.~Strutinsky, {\em Nucl. Phys. A} {\bf 95}  (1967) 420.

\bibitem{Strutinsky1968_NPA122-1}
V.~Strutinsky, {\em Nucl. Phys. A} {\bf 122}  (1968) 1.

\bibitem{Moeller1973_IAEA-SM-174-202}
P.~Moeller and J.~R. Nix, { Calculation of {{Fission Barriers}}}, in {\em
  Physics and {{Chemistry}} of {{Fission}}, {{Proceedings}} of the {{Third IAEA
  Symposium}} on {{Physics}} and {{Chemistry}} of {{Fission}}, {{Rochester}},
  {{New York}}, 13-17 {{August}} 1973\/},  ({International Atomic Energy
  Agency}, {Vienna, Austria}, April 1974), pp. 103--140.

\bibitem{Lu2012_PRC85-011301R}
B.-N. Lu, E.-G. Zhao and S.-G. Zhou, {\em Phys. Rev. C} {\bf 85}  (2012)
  011301(R).

\bibitem{Brack1972_RMP44-320}
M.~Brack, J.~Damgaard, A.~S. Jensen, H.~C. Pauli, V.~M. Strutinsky and C.~Y.
  Wong, {\em Rev. Mod. Phys.} {\bf 44}  (1972) 320.

\bibitem{Abusara2010_PRC82-044303}
H.~Abusara, A.~V. Afanasjev and P.~Ring, {\em Phys. Rev. C} {\bf 82}  (2010)
  044303.

\bibitem{Abusara2012_PRC85-024314}
H.~Abusara, A.~V. Afanasjev and P.~Ring, {\em Phys. Rev. C} {\bf 85}  (2012)
  024314.

\bibitem{Lu2014_PRC89-014323}
B.-N. Lu, J.~Zhao, E.-G. Zhao and S.-G. Zhou, {\em Phys. Rev. C} {\bf 89}
  (2014)   014323.

\bibitem{Schunck2016_RPP79-116301}
N.~Schunck and L.~M. Robledo, {\em Rept. Prog. Phys.} {\bf 79}  (2016)
  116301.

\bibitem{Schmidt2018_RPP81-106301}
K.-H. Schmidt and B.~Jurado, {\em Rept. Prog. Phys.} {\bf 81}  (2018)   106301.

\bibitem{Schunck2022_PPNP125-103963}
N.~Schunck and D.~Regnier, {\em Prog. Part. Nucl. Phys.} {\bf 125}  (2022)
  103963.

\bibitem{Sheikh2009_PRC80-011302R}
J.~A. Sheikh, W.~Nazarewicz and J.~C. Pei, {\em Phys. Rev. C} {\bf 80}  (2009)
   011302.

\bibitem{Kowal2010_PRC82-014303}
M.~Kowal, P.~Jachimowicz and A.~Sobiczewski, {\em Phys. Rev. C} {\bf 82}
  (2010)   014303.

\bibitem{Eichler2015_ApJ808-30}
M.~Eichler, A.~Arcones, A.~Kelic, O.~Korobkin, K.~Langanke, T.~Marketin,
  G.~{Martinez-Pinedo}, I.~Panov, T.~Rauscher, S.~Rosswog, C.~Winteler, N.~T.
  Zinner and F.-K. Thielemann, {\em Astrophys. J.} {\bf 808}  (2015)  ~30.

\bibitem{Sonzogni2016_PRL116-132502}
A.~A. Sonzogni, E.~A. McCutchan, T.~D. Johnson and P.~Dimitriou, {\em Phys.
  Rev. Lett.} {\bf 116}  (2016)   132502.

\bibitem{Agbemava2017_PRC95-054324}
S.~E. Agbemava, A.~V. Afanasjev, D.~Ray and P.~Ring, {\em Phys. Rev. C} {\bf
  95}  (2017)   054324.

\bibitem{Brown2018_NDS148-1}
D.~Brown, M.~Chadwick, R.~Capote, A.~Kahler, A.~Trkov, M.~Herman, A.~Sonzogni,
  Y.~Danon, A.~Carlson, M.~Dunn, D.~Smith, G.~Hale, G.~Arbanas, R.~Arcilla,
  C.~Bates, B.~Beck, B.~Becker, F.~Brown, R.~Casperson, J.~Conlin, D.~Cullen,
  M.-A. Descalle, R.~Firestone, T.~Gaines, K.~Guber, A.~Hawari, J.~Holmes,
  T.~Johnson, T.~Kawano, B.~Kiedrowski, A.~Koning, S.~Kopecky, L.~Leal,
  J.~Lestone, C.~Lubitz, J.~M{\'a}rquez~Dami{\'a}n, C.~Mattoon, E.~McCutchan,
  S.~Mughabghab, P.~Navratil, D.~Neudecker, G.~Nobre, G.~Noguere, M.~Paris,
  M.~Pigni, A.~Plompen, B.~Pritychenko, V.~Pronyaev, D.~Roubtsov, D.~Rochman,
  P.~Romano, P.~Schillebeeckx, S.~Simakov, M.~Sin, I.~Sirakov, B.~Sleaford,
  V.~Sobes, E.~Soukhovitskii, I.~Stetcu, P.~Talou, I.~Thompson, S.~{van der
  Marck}, L.~{Welser-Sherrill}, D.~Wiarda, M.~White, J.~Wormald, R.~Wright,
  M.~Zerkle, G.~{\v Z}erovnik and Y.~Zhu, {\em Nucl. Data Sheets} {\bf 148}
  (2018) 1.

\bibitem{Ge2020_EPJWoC239-09001}
Z.~Ge, R.~Xu, H.~Wu, Y.~Zhang, G.~Chen, Y.~Jin, N.~Shu, Y.~Chen, X.~Tao,
  Y.~Tian, P.~Liu, J.~Qian, J.~Wang, H.~Zhang, L.~Liu and X.~Huang, {\em EPJ
  Web Conf.} {\bf 239}  (2020)   09001.

\bibitem{Plompen2020_EPJA56-181}
A.~J.~M. Plompen, O.~Cabellos, C.~De~Saint~Jean, M.~Fleming, A.~Algora,
  M.~Angelone, P.~Archier, E.~Bauge, O.~Bersillon, A.~Blokhin, F.~Cantargi,
  A.~Chebboubi, C.~Diez, H.~Duarte, E.~Dupont, J.~Dyrda, B.~Erasmus,
  L.~Fiorito, U.~Fischer, D.~Flammini, D.~Foligno, M.~R. Gilbert, J.~R.
  Granada, W.~Haeck, F.-J. Hambsch, P.~Helgesson, S.~Hilaire, I.~Hill,
  M.~Hursin, R.~Ichou, R.~Jacqmin, B.~Jansky, C.~Jouanne, M.~A. Kellett, D.~H.
  Kim, H.~I. Kim, I.~Kodeli, A.~J. Koning, A.~Y. Konobeyev, S.~Kopecky, B.~Kos,
  A.~Kr{\'a}sa, L.~C. Leal, N.~Leclaire, P.~Leconte, Y.~O. Lee, H.~Leeb,
  O.~Litaize, M.~Majerle, J.~I. M{\'a}rquez~Dami{\'a}n, F.~{Michel-Sendis},
  R.~W. Mills, B.~Morillon, G.~Nogu{\`e}re, M.~Pecchia, S.~Pelloni,
  P.~Pereslavtsev, R.~J. Perry, D.~Rochman, A.~R{\"o}hrmoser, P.~Romain,
  P.~Romojaro, D.~Roubtsov, P.~Sauvan, P.~Schillebeeckx, K.~H. Schmidt,
  O.~Serot, S.~Simakov, I.~Sirakov, H.~Sj{\"o}strand, A.~Stankovskiy, J.~C.
  Sublet, P.~Tamagno, A.~Trkov, S.~{van der Marck}, F.~{\'A}lvarez-Velarde,
  R.~Villari, T.~C. Ware, K.~Yokoyama and G.~{\v Z}erovnik, {\em Eur. Phys. J.
  A} {\bf 56}  (2020)   181.

\bibitem{Zubov2002_PRC65-024308}
A.~Zubov, G.~Adamian, N.~Antonenko, S.~Ivanova and W.~Scheid, {\em Phys. Rev.
  C} {\bf 65}  (2002)   024308.

\bibitem{Xia2011_SciChinaPMA54S1-109}
C.~Xia, B.~Sun, E.~Zhao and S.~Zhou, {\em Sci. China-Phys. Mech. Astron.} {\bf
  54}  (2011) 109.

\bibitem{Wang2012_PRC85-041601R}
N.~Wang, E.-G. Zhao, W.~Scheid and S.-G. Zhou, {\em Phys. Rev. C} {\bf 85}
  (2012)   041601(R).

\bibitem{Adamian2021_EPJA57-89}
G.~G. Adamian, N.~V. Antonenko, H.~Lenske, L.~A. Malov and S.-G. Zhou, {\em
  Eur. Phys. J. A} {\bf 57}  (2021)  ~89.

\bibitem{Deng2023_PRC107-014616}
X.-Q. Deng and S.-G. Zhou, {\em Phys. Rev. C} {\bf 107}  (2023)   014616.

\bibitem{Howard1980_ADNDT25-219}
W.~Howard and P.~M{\"o}ller, {\em Atom. Data Nucl. Data Tables} {\bf 25}
  (1980) 219.

\bibitem{Myers1996_NPA601-141}
W.~Myers and W.~Swiatecki, {\em Nucl. Phys. A} {\bf 601}  (1996) 141.

\bibitem{Jachimowicz2012_PRC85-034305}
P.~Jachimowicz, M.~Kowal and J.~Skalski, {\em Phys. Rev. C} {\bf 85}  (2012)
  034305.

\bibitem{Staszczak2013_PRC87-024320}
A.~Staszczak, A.~Baran and W.~Nazarewicz, {\em Phys. Rev. C} {\bf 87}  (2013)
  024320.

\bibitem{Pomorski2013_PST154-014023}
K.~Pomorski, {\em Phys. Scr.} {\bf T154}  (2013)   014023.

\bibitem{Zhong2014_CTP62-405}
C.-L. Zhong and T.-S. Fan, {\em Commun. Theor. Phys.} {\bf 62}  (2014) 405.

\bibitem{Moeller2015_PRC91-024310}
P.~M{\"o}ller, A.~J. Sierk, T.~Ichikawa, A.~Iwamoto and M.~Mumpower, {\em Phys.
  Rev. C} {\bf 91}  (2015)   024310.

\bibitem{Pomorski2017_PS92-064006}
K.~Pomorski, B.~{Nerlo-Pomorska} and J.~Bartel, {\em Phys. Scr.} {\bf 92}
  (2017)   064006.

\bibitem{Chai2018_ChinPhysC42-054101}
Q.-Z. Chai, W.-J. Zhao, M.-L. Liu and H.-L. Wang, {\em Chin. Phys. C} {\bf 42}
  (2018)   054101.

\bibitem{Jachimowicz2020_PRC101-014311}
P.~Jachimowicz, M.~Kowal and J.~Skalski, {\em Phys. Rev. C} {\bf 101}  (2020)
  014311.

\bibitem{Zhu2020_CTP72-105301}
X.~Zhu, Z.-M. Wang, W.-J. Zhu, C.-L. Zhong, Y.-M. Zhang, L.-Y. Liao and T.-S.
  Fan, {\em Commun. Theor. Phys.} {\bf 72}  (2020)   105301.

\bibitem{Mamdouh2001_NPA679-337}
A.~Mamdouh, J.~Pearson, M.~Rayet and F.~Tondeur, {\em Nucl. Phys. A} {\bf 679}
  (2001) 337.

\bibitem{Delaroche2006_NPA771-103}
J.-P. Delaroche, M.~Girod, H.~Goutte and J.~Libert, {\em Nucl. Phys. A} {\bf
  771}  (2006) 103.

\bibitem{Goriely2009_PRC79-024612}
S.~Goriely, S.~Hilaire, A.~J. Koning, M.~Sin and R.~Capote, {\em Phys. Rev. C}
  {\bf 79}  (2009)   024612.

\bibitem{Giuliani2013_PRC88-054325}
S.~A. Giuliani and L.~M. Robledo, {\em Phys. Rev. C} {\bf 88}  (2013)   054325.

\bibitem{Rodriguez-Guzman2014_PRC89-054310}
R.~{Rodr{\'i}guez-Guzm{\'a}n} and L.~M. Robledo, {\em Phys. Rev. C} {\bf 89}
  (2014)   054310.

\bibitem{Zhou2016_PS91-063008}
S.-G. Zhou, {\em Phys. Scripta} {\bf 91}  (2016)   063008.

\bibitem{Giuliani2018_PRC97-034323}
S.~A. Giuliani, G.~{Mart{\'i}nez-Pinedo} and L.~M. Robledo, {\em Phys. Rev. C}
  {\bf 97}  (2018)   034323.

\bibitem{Ling2020_EPJA56-180}
C.~Ling, C.~Zhou and Y.~Shi, {\em Eur. Phys. J. A} {\bf 56}  (2020)   180.

\bibitem{Taninah2020_PRC102-054330}
A.~Taninah, S.~E. Agbemava and A.~V. Afanasjev, {\em Phys. Rev. C} {\bf 102}
  (2020)   054330.

\bibitem{Serot1986_ANP16-1}
B.~D. Serot and J.~D. Walecka, {\em The Relativistic Nuclear Many-Body
  Problem}, no.~16 in Advances in {{Nuclear Physics}} ({Plenum Press}, {New
  York}, 1986).

\bibitem{Reinhard1989_RPP52-439}
P.~G. Reinhard, {\em Rept. Prog. Phys.} {\bf 52}  (1989)   439.

\bibitem{Ring1996_PPNP37-193}
P.~Ring, {\em Prog. Part. Nucl. Phys.} {\bf 37}  (1996) 193.

\bibitem{Vretenar2005_PR409-101}
D.~Vretenar, A.~Afanasjev, G.~Lalazissis and P.~Ring, {\em Phys. Rept.} {\bf
  409}  (2005) 101.

\bibitem{Meng2006_PPNP57-470}
J.~Meng, H.~Toki, S.~Zhou, S.~Zhang, W.~Long and L.~Geng, {\em Prog. Part.
  Nucl. Phys.} {\bf 57}  (2006) 470.

\bibitem{Niksic2011_PPNP66-519}
T.~Nik{\v s}i{\'c}, D.~Vretenar and P.~Ring, {\em Prog. Part. Nucl. Phys.} {\bf
  66}  (2011) 519.

\bibitem{Meng2013_FrontPhys8-55}
J.~Meng, J.~Peng, S.-Q. Zhang and P.-W. Zhao, {\em Front. Phys.} {\bf 8}
  (2013) 55.

\bibitem{Liang2015_PR570-1}
H.~Liang, J.~Meng and S.-G. Zhou, {\em Phys. Rept.} {\bf 570}  (2015) 1.

\bibitem{Meng2015_JPG42-093101}
J.~Meng and S.~G. Zhou, {\em J. Phys. G: Nucl. Part. Phys.} {\bf 42}  (2015)
  093101.

\bibitem{Meng2016_RDFNS}
J.~Meng, {\em Relativistic {{Density Functional}} for {{Nuclear Structure}}},
  no.~10 in International {{Review}} of {{Nuclear Physics}} ({World
  Scientific}, {Singapore}, March 2016).

\bibitem{Shen2019_PPNP109-103713}
S.~Shen, H.~Liang, W.~H. Long, J.~Meng and P.~Ring, {\em Prog. Part. Nucl.
  Phys.} {\bf 109}  (2019)   103713.

\bibitem{Geng2007_CPL24-1865}
L.-S. Geng, M.~Jie and T.~Hiroshi, {\em Chin. Phys. Lett.} {\bf 24}  (2007)
  1865.

\bibitem{Zhang2010_PRC81-034302}
W.~Zhang, Z.~P. Li, S.~Q. Zhang and J.~Meng, {\em Phys. Rev. C} {\bf 81}
  (2010)   034302.

\bibitem{Wei2010_ChinPhysC34-1094}
Z.~Wei, L.~{Zhi-Pan} and Z.~{Shuang-Quan}, {\em Chin. Phys. C} {\bf 34}  (2010)
  1094.

\bibitem{Guo2010_PRC82-047301}
J.-Y. Guo, P.~Jiao and X.-Z. Fang, {\em Phys. Rev. C} {\bf 82}  (2010)
  047301.

\bibitem{Qiu2022_PRC106-034301}
Y.-T. Qiu, X.-W. Wang and J.-Y. Guo, {\em Phys. Rev. C} {\bf 106}  (2022)
  034301.

\bibitem{Gambhir1990_APNY198-132}
Y.~Gambhir, P.~Ring and A.~Thimet, {\em Annals Phys.} {\bf 198}  (1990) 132.

\bibitem{Meng2020_SciChinaPMA63-212011}
X.~Meng, B.~Lu and S.~Zhou, {\em Sci. China-Phys. Mech. Astron.} {\bf 63}
  (2020)   212011.

\bibitem{Wang2022_ChinPhysC46-024107}
X.-Q. Wang, X.-X. Sun and S.-G. Zhou, {\em Chin. Phys. C} {\bf 46}  (2022)
  024107.

\bibitem{Zhao2012_PRC86-057304}
J.~Zhao, B.-N. Lu, E.-G. Zhao and S.-G. Zhou, {\em Phys. Rev. C} {\bf 86}
  (2012)   057304.

\bibitem{Xu2022_PLB833-137287}
W.~Xu, S.~Wang, C.~Liu, X.~Wu, R.~Guo, B.~Qi, J.~Zhao, A.~Rohilla, H.~Jia,
  G.~Li, Y.~Zheng, C.~Li, X.~Han, L.~Mu, X.~Xiao, S.~Wang, D.~Sun, Z.~Li,
  Y.~Zhang, C.~Wang and Y.~Li, {\em Phys. Lett. B} {\bf 833}  (2022)   137287.

\bibitem{Liu2016_PRL116-112501}
C.~Liu, S.~Y. Wang, R.~A. Bark, S.~Q. Zhang, J.~Meng, B.~Qi, P.~Jones, S.~M.
  Wyngaardt, J.~Zhao, C.~Xu, S.-G. Zhou, S.~Wang, D.~P. Sun, L.~Liu, Z.~Q. Li,
  N.~B. Zhang, H.~Jia, X.~Q. Li, H.~Hua, Q.~B. Chen, Z.~G. Xiao, H.~J. Li,
  L.~H. Zhu, T.~D. Bucher, T.~Dinoko, J.~Easton, K.~Juh{\'a}sz, A.~Kamblawe,
  E.~Khaleel, N.~Khumalo, E.~A. Lawrie, J.~J. Lawrie, S.~N.~T. Majola, S.~M.
  Mullins, S.~Murray, J.~Ndayishimye, D.~Negi, S.~P. Noncolela, S.~S.
  Ntshangase, B.~M. Nyak{\'o}, J.~N. Orce, P.~Papka, J.~F. {Sharpey-Schafer},
  O.~Shirinda, P.~Sithole, M.~A. Stankiewicz and M.~Wiedeking, {\em Phys. Rev.
  Lett.} {\bf 116}  (2016)   112501.

\bibitem{Wang2019_PLB792-454}
Y.~Wang, S.~Zhang, P.~Zhao and J.~Meng, {\em Phys. Lett. B} {\bf 792}  (2019)
  454.

\bibitem{Chen2016_PRC94-021301R}
X.~C. Chen, J.~Zhao, C.~Xu, H.~Hua, T.~M. Shneidman, S.~G. Zhou, X.~G. Wu,
  X.~Q. Li, S.~Q. Zhang, Z.~H. Li, W.~Y. Liang, J.~Meng, F.~R. Xu, B.~Qi, Y.~L.
  Ye, D.~X. Jiang, Y.~Y. Cheng, C.~He, J.~J. Sun, R.~Han, C.~Y. Niu, C.~G. Li,
  P.~J. Li, C.~G. Wang, H.~Y. Wu, Z.~H. Li, H.~Zhou, S.~P. Hu, H.~Q. Zhang,
  G.~S. Li, C.~Y. He, Y.~Zheng, C.~B. Li, H.~W. Li, Y.~H. Wu, P.~W. Luo and
  J.~Zhong, {\em Phys. Rev. C} {\bf 94}  (2016)   021301.

\bibitem{Wang2022_PRC105-044316}
Y.~Y. Wang, Q.~B. Chen and S.~Q. Zhang, {\em Phys. Rev. C} {\bf 105}  (2022)
  044316.

\bibitem{Wang2022_PRC106-L011303}
C.~G. Wang, R.~Han, C.~Xu, H.~Hua, R.~A. Bark, S.~Q. Zhang, S.~Y. Wang, T.~M.
  Shneidman, S.~G. Zhou, J.~Meng, S.~M. Wyngaardt, A.~C. Dai, F.~R. Xu, X.~Q.
  Li, Z.~H. Li, Y.~L. Ye, D.~X. Jiang, C.~G. Li, C.~Y. Niu, Z.~Q. Chen, H.~Y.
  Wu, D.~W. Luo, S.~Wang, D.~P. Sun, C.~Liu, Z.~Q. Li, N.~B. Zhang, R.~J. Guo,
  P.~Jones, E.~A. Lawrie, J.~J. Lawrie, J.~F. {Sharpey-Schafer}, M.~Wiedeking,
  S.~N.~T. Majola, T.~D. Bucher, T.~Dinoko, B.~Maqabuka, L.~Makhathini,
  L.~Mdletshe, O.~Shirinda and K.~Sowazi, {\em Phys. Rev. C} {\bf 106}  (2022)
   L011303.

\bibitem{Lu2011_PRC84-014328}
B.-N. Lu, E.-G. Zhao and S.-G. Zhou, {\em Phys. Rev. C} {\bf 84}  (2011)
  014328.

\bibitem{Lu2014_PRC89-044307}
B.-N. Lu, E.~Hiyama, H.~Sagawa and S.-G. Zhou, {\em Phys. Rev. C} {\bf 89}
  (2014)   044307.

\bibitem{Rong2020_PLB807-135533}
Y.-T. Rong, P.~Zhao and S.-G. Zhou, {\em Phys. Lett. B} {\bf 807}  (2020)
  135533.

\bibitem{Rong2021_PRC104-054321}
Y.-T. Rong, Z.-H. Tu and S.-G. Zhou, {\em Phys. Rev. C} {\bf 104}  (2021)
  054321.

\bibitem{Chen2021_SciChinaPMA64-282011}
C.~Chen, Q.-K. Sun, Y.-X. Li and T.-T. Sun, {\em Sci. China-Phys. Mech.
  Astron.} {\bf 64}  (2021)   282011.

\bibitem{Sun2022_ChinPhysC46-074106}
Q.-K. Sun, T.-T. Sun, W.~Zhang, S.-S. Zhang and C.~Chen, {\em Chin. Phys. C}
  {\bf 46}  (2022)   074106.

\bibitem{Zhao2015_PRC92-064315}
J.~Zhao, B.-N. Lu, T.~Nik{\v s}i{\'c} and D.~Vretenar, {\em Phys. Rev. C} {\bf
  92}  (2015)   064315.

\bibitem{Zhao2016_PRC93-044315}
J.~Zhao, B.-N. Lu, T.~Nik{\v s}i{\'c}, D.~Vretenar and S.-G. Zhou, {\em Phys.
  Rev. C} {\bf 93}  (2016)   044315.

\bibitem{Zhao2019_PRC99-014618}
J.~Zhao, T.~Nik{\v s}i{\'c}, D.~Vretenar and S.-G. Zhou, {\em Phys. Rev. C}
  {\bf 99}  (2019)   014618.

\bibitem{Zhao2019_PRC99-054613}
J.~Zhao, J.~Xiang, Z.-P. Li, T.~Nik{\v s}i{\'c}, D.~Vretenar and S.-G. Zhou,
  {\em Phys. Rev. C} {\bf 99}  (2019)   054613.

\bibitem{Zhao2020_PRC102-054606}
J.~Zhao, T.~Nik{\v s}i{\'c} and D.~Vretenar, {\em Phys. Rev. C} {\bf 102}
  (2020)   054606.

\bibitem{Zhao2020_PRC101-064605}
J.~Zhao, T.~Nik{\v s}i{\'c}, D.~Vretenar and S.-G. Zhou, {\em Phys. Rev. C}
  {\bf 101}  (2020)   064605.

\bibitem{Zhao2021_PRC104-044612}
J.~Zhao, T.~Nik{\v s}i{\'c} and D.~Vretenar, {\em Phys. Rev. C} {\bf 104}
  (2021)   044612.

\bibitem{Ren2022_PRC105-044313}
Z.~X. Ren, J.~Zhao, D.~Vretenar, T.~Nik{\v s}i{\'c}, P.~W. Zhao and J.~Meng,
  {\em Phys. Rev. C} {\bf 105}  (2022)   044313.

\bibitem{Zhao2022_PRC105-054604}
J.~Zhao, T.~Nik{\v s}i{\'c} and D.~Vretenar, {\em Phys. Rev. C} {\bf 105}
  (2022)   054604.

\bibitem{Zhao2022_PRC106-054609}
J.~Zhao, T.~Nik{\v s}i{\'c} and D.~Vretenar, {\em Phys. Rev. C} {\bf 106}
  (2022)   054609.

\bibitem{Mercier2020_PRC102-011301}
F.~Mercier, J.~Zhao, R.-D. Lasseri, J.-P. Ebran, E.~Khan, T.~Nik{\v s}i{\'c}
  and D.~Vretenar, {\em Phys. Rev. C} {\bf 102}  (2020)   011301.

\bibitem{Mercier2021_PRL127-012501}
F.~Mercier, J.~Zhao, J.-P. Ebran, E.~Khan, T.~Nik{\v s}i{\'c} and D.~Vretenar,
  {\em Phys. Rev. Lett.} {\bf 127}  (2021)   012501.

\bibitem{Rong2023_PLB-submitted}
Y.-T. Rong, X.-Y. Wu, B.-N. Lu and J.-M. Yao, Anatomy of octupole correlations
  in {\textsuperscript{96}}{{Zr}} with a symmetry-restored
  multidimensionally-constrained covariant density functional theory,
  submitted to {\em Phys. Lett. B}.

\bibitem{Zhao2010_PRC82-054319}
P.~W. Zhao, Z.~P. Li, J.~M. Yao and J.~Meng, {\em Phys. Rev. C} {\bf 82}
  (2010)   054319.

\bibitem{Tian2006_CPL23-3226}
Y.~Tian and Z.-y. Ma, {\em Chin. Phys. Lett.} {\bf 23}  (2006) 3226.

\bibitem{Tian2009_PLB676-44}
Y.~Tian, Z.~Ma and P.~Ring, {\em Phys. Lett. B} {\bf 676}  (2009) 44.

\bibitem{Tian2009_PRC79-064301}
Y.~Tian, Z.-y. Ma and P.~Ring, {\em Phys. Rev. C} {\bf 79}  (2009)   064301.

\bibitem{Zhang2021_PRL126-152502}
Z.~Y. Zhang, H.~B. Yang, M.~H. Huang, Z.~G. Gan, C.~X. Yuan, C.~Qi, A.~N.
  Andreyev, M.~L. Liu, L.~Ma, M.~M. Zhang, Y.~L. Tian, Y.~S. Wang, J.~G. Wang,
  C.~L. Yang, G.~S. Li, Y.~H. Qiang, W.~Q. Yang, R.~F. Chen, H.~B. Zhang, Z.~W.
  Lu, X.~X. Xu, L.~M. Duan, H.~R. Yang, W.~X. Huang, Z.~Liu, X.~H. Zhou, Y.~H.
  Zhang, H.~S. Xu, N.~Wang, H.~B. Zhou, X.~J. Wen, S.~Huang, W.~Hua, L.~Zhu,
  X.~Wang, Y.~C. Mao, X.~T. He, S.~Y. Wang, W.~Z. Xu, H.~W. Li, Z.~Z. Ren and
  S.~G. Zhou, {\em Phys. Rev. Lett.} {\bf 126}  (2021)   152502.

\bibitem{Huang2021_ChinPhysC45-030002}
W.~Huang, M.~Wang, F.~Kondev, G.~Audi and S.~Naimi, {\em Chin. Phys. C} {\bf
  45}  (2021)   030002.

\bibitem{Wang2021_ChinPhysC45-030003}
M.~Wang, W.~Huang, F.~Kondev, G.~Audi and S.~Naimi, {\em Chin. Phys. C} {\bf
  45}  (2021)   030003.

\bibitem{Zhang2022_ADNDT144-101488}
K.~Zhang, M.-K. Cheoun, Y.-B. Choi, P.~S. Chong, J.~Dong, Z.~Dong, X.~Du,
  L.~Geng, E.~Ha, X.-T. He, C.~Heo, M.~C. Ho, E.~J. In, S.~Kim, Y.~Kim, C.-H.
  Lee, J.~Lee, H.~Li, Z.~Li, T.~Luo, J.~Meng, M.-H. Mun, Z.~Niu, C.~Pan,
  P.~Papakonstantinou, X.~Shang, C.~Shen, G.~Shen, W.~Sun, X.-X. Sun, C.~K.
  Tam, {Thaivayongnou}, C.~Wang, X.~Wang, S.~H. Wong, J.~Wu, X.~Wu, X.~Xia,
  Y.~Yan, R.~W.-Y. Yeung, T.~C. Yiu, S.~Zhang, W.~Zhang, X.~Zhang, Q.~Zhao and
  S.-G. Zhou, {\em At. Data Nucl. Data Tables} {\bf 144}  (2022)   101488.

\bibitem{Angeli2013_ADNDT99-69}
I.~Angeli and K.~Marinova, {\em Atom. Data Nucl. Data Tables} {\bf 99}  (2013)
  69.

\bibitem{Zhou2010_PRC82-011301R}
S.-G. Zhou, J.~Meng, P.~Ring and E.-G. Zhao, {\em Phys. Rev. C} {\bf 82}
  (2010)   011301.

\bibitem{Li2012_PRC85-024312}
L.~Li, J.~Meng, P.~Ring, E.-G. Zhao and S.-G. Zhou, {\em Phys. Rev. C} {\bf 85}
   (2012)   024312.

\bibitem{Chen2012_PRC85-067301}
Y.~Chen, L.~Li, H.~Liang and J.~Meng, {\em Phys. Rev. C} {\bf 85}  (2012)
  067301.

\bibitem{Capote2009_NDS110-3107}
R.~Capote, M.~Herman, P.~Oblo{\v z}insk{\'y}, P.~Young, S.~Goriely, T.~Belgya,
  A.~Ignatyuk, A.~Koning, S.~Hilaire, V.~Plujko, M.~Avrigeanu, O.~Bersillon,
  M.~Chadwick, T.~Fukahori, Z.~Ge, Y.~Han, S.~Kailas, J.~Kopecky, V.~Maslov,
  G.~Reffo, M.~Sin, E.~Soukhovitskii and P.~Talou, {\em Nucl. Data Sheets} {\bf
  110}  (2009) 3107.

\bibitem{Zhang1993_NSE114-55}
J.~Zhang, {\em Nucl. Sci. Eng.} {\bf 114}  (1993) 55.

\bibitem{Zhang2012_FUNF-Manual}
J.~Zhang, {\em {User Manual of FUNF Code}} ({Atomic Energy Press}, {Beijing},
  2012).

\bibitem{Sun2022_PriCom}
X.-J. Sun and J.-S. Zhang, Private {{Communication}}  (2022).

\bibitem{Cwiok1994_PLB322-304}
S.~{\'C}wiok, W.~Nazarewicz, J.~Saladin, W.~P{\l}{\'o}ciennik and A.~Johnson,
  {\em Phys. Lett. B} {\bf 322}  (1994) 304.

\bibitem{Rutz1995_NPA590-680}
K.~Rutz, J.~Maruhn, P.-G. Reinhard and W.~Greiner, {\em Nucl. Phys. A} {\bf
  590}  (1995) 680.

\bibitem{Krasznahorkay1998_APHA7-35}
A.~Krasznahorkay, M.~Hunyadi, M.~Csatl{\'o}s, A.~Ba{\l}anda, A.~Gollwitzer,
  G.~Graw, J.~Guly{\'a}s, D.~Habs, R.~Hertenberger, Z.~M{\'a}t{\'e}, H.~J.
  Maier, D.~Rudolph, P.~Thirolf, J.~Tim{\'a}r and B.~D. Valnion, {\em Acta
  Phys. Hung. A} {\bf 7}  (1998) 35.

\bibitem{Marinov2001_IJMPE10-209}
A.~Marinov, S.~Gelberg, D.~Kolb and J.~L. Weil, {\em Int. J. Mod. Phys. E} {\bf
  10}  (2001) 209.

\bibitem{Thirolf2001_APHA13-111}
P.~G. Thirolf, S.~Y. {van der Werf}, J.~Ott, Z.~M{\'a}t{\'e}, J.~Guly{\'a}s,
  M.~Csatl{\'o}s, M.~Hunyadi, A.~Krasznahorkay, T.~Faestermann, A.~Metz, H.~J.
  Maier, R.~Hertenberger, G.~Graw, Y.~Eisermann, D.~Habs and D.~Gassmann, {\em
  Acta Phys. Hung. A} {\bf 13}  (2001) 111.

\bibitem{Thirolf2002_PPNP49-325}
P.~Thirolf and D.~Habs, {\em Prog. Part. Nucl. Phys.} {\bf 49}  (2002) 325.

\bibitem{Krasznahorkay2003_APHA18-323}
A.~Krasznahorkay, M.~Csatl{\'o}s, L.~Csige, Y.~Eisermann, T.~Faestermann,
  Z.~G{\'a}csi, G.~Graw, J.~Guly{\'a}s, D.~Habs, M.~N. Harakeh, M.~Heil,
  R.~Hertenberger, F.~Kappeler, A.~J. Krasznahorkay, H.~J. Maier, Z.~Mate,
  R.~Reifarth, P.~G. Thirolf, J.~Tim{\'a}r and H.~F. Wirth, {\em Acta Phys.
  Hung. A} {\bf 18}  (2003) 323.

\bibitem{Sin2006_PRC74-014608}
M.~Sin, R.~Capote, A.~Ventura, M.~Herman and P.~Oblo{\v z}insk{\'Y}, {\em Phys.
  Rev. C} {\bf 74}  (2006)   014608.

\bibitem{Kowal2012_PRC85-061302R}
M.~Kowal and J.~Skalski, {\em Phys. Rev. C} {\bf 85}  (2012)   061302.

\bibitem{Ichikawa2013_PRC87-054326}
T.~Ichikawa, P.~M{\"o}ller and A.~J. Sierk, {\em Phys. Rev. C} {\bf 87}  (2013)
    054326.

\bibitem{Jachimowicz2013_PRC87-044308}
P.~Jachimowicz, M.~Kowal and J.~Skalski, {\em Phys. Rev. C} {\bf 87}  (2013)
  044308.

\bibitem{McDonnell2013_PRC87-054327}
J.~D. McDonnell, W.~Nazarewicz and J.~A. Sheikh, {\em Phys. Rev. C} {\bf 87}
  (2013)   054327.

\end{thebibliography}

\end{document}